\documentclass[onecolumn]{aastex63}
\usepackage{float,graphicx,amsmath,multirow}
\usepackage{color}
\usepackage[version=4]{mhchem}
\usepackage{booktabs}

\usepackage{xcolor}
\usepackage{xifthen}
\usepackage[normalem]{ulem}
\newcommand{\stkout}[1]{\ifmmode\text{\sout{\ensuremath{#1}}}\else\sout{#1}\fi}
\newcommand{\edited}[2]{\ifthenelse{\isempty{#1}}{\textbf{#2}}{\ifthenelse{\isempty{#2}}{\textcolor{gray}{\stkout{#1}}}{\textcolor{gray}{\stkout{#1}} \textbf{#2}}}}

\usepackage{xstring,xifthen}
\newcommand{\replacedecimal}[2]{\ifthenelse{\isin{.}{#1}}{\text{\StrBefore{#1}{.}}\ensuremath{\overset{#2}{.}}\text{\StrBehind{#1}{.}}}{#1\ensuremath{^{#2}}}}
\newcommand{\hms}[3]{\ensuremath{#1\overset{\text{h}}{\phantom{.}}#2\overset{\text{m}}{\phantom{.}}\replacedecimal{#3}{\text{s}}}}
\newcommand{\dms}[3]{\ensuremath{#1\overset{\circ}{\phantom{.}}#2\overset{\prime}{\phantom{.}}\replacedecimal{#3}{\prime\prime}}}

\begin{document}

\title{Detection of Interstellar \ce{HC4NC} and an Investigation of Isocyanopolyyne Chemistry in TMC-1 Conditions}
\author{Ci Xue}
\affiliation{Department of Chemistry, University of Virginia, Charlottesville, VA 22904, USA}
\author{Eric R. Willis}
\affiliation{Department of Chemistry, University of Virginia, Charlottesville, VA 22904, USA}
\author{Ryan A. Loomis}
\affiliation{National Radio Astronomy Observatory, Charlottesville, VA 22903, USA}
\author{Kin Long Kelvin Lee}
\affiliation{Center for Astrophysics $\mid$ Harvard~\&~Smithsonian, Cambridge, MA 02138, USA}
\author{Andrew M. Burkhardt}
\affiliation{Center for Astrophysics $\mid$ Harvard~\&~Smithsonian, Cambridge, MA 02138, USA}
\author{Christopher N. Shingledecker}
\affiliation{Department of Physics and Astronomy, Benedictine College, Atchison, KS 66002, USA}
\affiliation{Center for Astrochemical Studies, Max Planck Intitute for Extraterrestrial Physics, Garching, Germany}
\affiliation{Institute for Theoretical Chemistry, University of Stuttgart, Stuttgart, Germany}
\author{Steven B. Charnley}
\affiliation{Astrochemistry Laboratory and the Goddard Center for Astrobiology, NASA Goddard Space Flight Center, Greenbelt, MD 20771, USA}
\author{Martin A. Cordiner}
\affiliation{Astrochemistry Laboratory and the Goddard Center for Astrobiology, NASA Goddard Space Flight Center, Greenbelt, MD 20771, USA}
\affiliation{Institute for Astrophysics and Computational Sciences, The Catholic University of America, Washington, DC 20064, USA}
\author{Sergei Kalenskii}
\affiliation{Astro Space Center, Lebedev Physical Institute, Russian Academy of Sciences, Moscow, Russia}
\author{Michael C. McCarthy}
\affiliation{Center for Astrophysics $\mid$ Harvard~\&~Smithsonian, Cambridge, MA 02138, USA}
\author{Eric Herbst}
\affiliation{Department of Chemistry, University of Virginia, Charlottesville, VA 22904, USA}
\affiliation{Department of Astronomy, University of Virginia, Charlottesville, VA 22904, USA}
\author{Anthony J. Remijan}
\affiliation{National Radio Astronomy Observatory, Charlottesville, VA 22903, USA}
\author{Brett A. McGuire}
\affiliation{Department of Chemistry, Massachusetts Institute of Technology, Cambridge, MA 02139, USA}
\affiliation{National Radio Astronomy Observatory, Charlottesville, VA 22903, USA}
\affiliation{Center for Astrophysics $\mid$ Harvard~\&~Smithsonian, Cambridge, MA 02138, USA}

\correspondingauthor{Ci Xue, Brett A. McGuire}
\email{cx5up@virginia.edu, brettmc@mit.edu}

\begin{abstract}
We report an astronomical detection of \ce{HC4NC} for the first time in the interstellar medium with the Green Bank Telescope toward the TMC-1 molecular cloud with a minimum significance of $10.5 \sigma$. The total column density and excitation temperature of \ce{HC4NC} are determined to be $3.29^{+8.60}_{-1.20}\times 10^{11}$~cm$^{-2}$ and $6.7^{+0.3}_{-0.3} \mathrm{\ K}$, respectively, using the MCMC analysis. In addition to \ce{HC4NC}, \ce{HCCNC} is distinctly detected whereas no clear detection of \ce{HC6NC} is made. We propose that the dissociative recombination of the protonated cyanopolyyne, \ce{HC5NH+}, and the protonated isocyanopolyyne, \ce{HC4NCH+}, are the main formation mechanisms for \ce{HC4NC} while its destruction is dominated by reactions with simple ions and atomic carbon. With the proposed chemical networks, the observed abundances of \ce{HC4NC} and \ce{HCCNC} are reproduced satisfactorily.
\end{abstract}

\section{Introduction\label{sec:intro}}
Understanding the formation and destruction routes of molecules in astronomical environments remains one of the challenging issues in modern astrochemistry. Increasingly sensitive astronomical observations can reveal detailed information about the chemical inventories present in interstellar sources. Laboratory experiments and astrochemical modelling can then work in tandem to uncover the chemical mechanisms underlying these molecular inventories. However, there are still deficiencies in our understanding of the chemistry of interstellar sources. For example, in spite of proposed formation routes through grain chemistry, gas-phase formation routes cannot be ruled out as a viable pathway for the formation of large astronomical molecules (LAMs) \citep[and references therein]{2015MNRAS.449L..16B, 2017MNRAS.467..737C, 2017ApJ...850..105A}. The question remains as to how to better model the chemistry present in these astronomical environments and make the models more predictive. In turn, these robust models could then suggest further chemical species to be investigated both in the laboratory and through astronomical observations.

Structural isomers are a promising class of molecules for improving the accuracy of models. Structural isomers contain the same constituent atoms but are arranged in different elemental configurations \citep{2019ApJ...871..112X}. One of the most well studied isomeric pairs in astronomical environments is that of hydrogen cyanide (\ce{HCN}) and isocyanide (\ce{HNC}) \citep{1992A&A...256..595S,1997ApJ...483..235T,1998ApJ...503..717H,2000MNRAS.311..869H,2006A&A...456.1037T,2015ApJ...807L..15G}. At 100 K, under thermal equilibrium conditions, the relative abundance ratio between \ce{HNC} and \ce{HCN} is $\sim$10$^{-30}$ \citep{1977Natur.270...39B}. However, it is well known that under dark cloud conditions, such as those found in the Taurus Molecular Cloud 1 (TMC-1), the abundance ratio approaches $\sim$1 \citep{1984ApJ...282..516I}, indicating that thermodynamic equilibrium certainly does not apply to the two species in these regions \citep{1989ApJ...347..855B}. Instead, measured column densities toward these sources are dominated by the kinetics of chemical reactions in the gas phase; these measurements give observational constraints on the chemical formation and destruction networks \citep{2014ApJ...787...74G}. As such, measuring the relative abundance ratios for pairs of chemical isomers, and incorporating isomer-specific chemistry into chemical networks, can be a powerful tool to improving the predictive power of these models. Here, we focus on exploiting the cyanide and isocyanide pairs of isomers.

The family of astronomically-detected cyanides includes \ce{HCN}, methyl cyanide (\ce{CH3CN}), vinyl cyanide (\ce{CH2CHCN}), ethyl cyanide (\ce{CH3CH2CN}), and other species including isocyanogen (\ce{CNCN}), E-cyanomethanimine (\ce{E-HNCHCN}), glycolonitrile (\ce{HOCH2CN}) and many others \citep[and references therein]{2018ApJS..239...17M,2019MNRAS.484L..43Z}. Some of these species are found in high abundance and are readily detectable in a variety of interstellar environments \citep{1997ApJ...480L..67M,2005ApJS..157..279A,2014A&A...572A..44L,2019ApJ...872...61H}. In contrast to the numerous detection of cyanides in astronomical environments, there have been very few confirmed detection of isocyanides, such as methyl isocyanide (\ce{CH3NC}) \citep{2005ApJ...632..333R, 2013A&A...557A.101G}. Most recently, the Protostellar Interferometric Line Survey (PILS) observed \ce{CH3NC} in a solar-type star, IRAS 16293-2422, for the first time toward a source of this type \citep{2018AnA...617A..95C}. Despite that, there have been no successful detections of \ce{CH2CHCH2NC} \citep{2013ApJ...777..120H} or \ce{CH3CH2NC} \citep{2005ApJ...632..333R, 2018A&A...610A..44M}.

Alongside \ce{CH3CN}, one of the most frequently observed families of cyanide species, especially in cold sources, are the cyanopolyynes (\ce{HC_{2n}CN}) \citep{1978ApJ...223L.105B,1978MNRAS.183P..45L,1998ApJ...508..286B}. Yet, despite their relative ubiquity, the only isocyanide version that has been successfully detected is \ce{HCCNC} \citep{1992ApJ...386L..51K}, the isomer of \ce{HC3N}. \citet{2005ApJ...632..333R} first searched for isocyanodiacetylene (\ce{HC4NC}), the isomer of \ce{HC5N}, toward Sagittarius B2(N). To the best of our knowledge, this has been the only attempt to detect this molecule in astronomical environments, setting an upper limit on the abundance ratio to \ce{HC5N} as $0.03$. In this work, we report the first astronomical detection of \ce{HC4NC} using the data available from the GOTHAM (Green Bank Telescope Observations of TMC-1: Hunting for Aromatic Molecules) observational program of TMC-1 \citep{McGuire:2020bb}. The detection of \ce{HC4NC} along with new observations of \ce{HCCNC} and an upper limit to the abundance of \ce{HC6NC}, have been used to better constrain the gas-phase formation models of both cyanopolyynes and isocyanopolyynes (\ce{HC_{2n}NC}) under TMC-1 conditions. {The interplay between –CN and –NC formation chemistry can also provide insights into the physical conditions and history of the sources where these species are detected, therefore making new mechanistic insights into -CN vs -NC chemistry particularly relevant for both new and continuing problems such as the \ce{HCN}/\ce{HNC} abundance ratio \citep[e.g.][]{2020A&A...635A...4H}.} 

In \S\ref{sec:spec}, we describe the molecular properties of \ce{HC4NC}. \S\ref{sec:results} presents the detection of \ce{HC4NC} with the GOTHAM observations and the observational analyses. The results of the analyses are used to constrain the new chemical formation network developed to account for the formation of \ce{HC4NC} in \S\ref{sec:discussion}. Finally, in \S\ref{sec:conclusions}, we summarize our results and describe the next steps in refining the chemical network and searches for larger isocyanopolyynes towards other astronomical sources.
%discuss the level of agreement

\section{Spectroscopic Properties \label{sec:spec}}
The \ce{HC4NC} molecule has a linear equilibrium structure \citep{2006CPL...428..245G}. For this work, transition frequencies of \ce{HC4NC} were taken from the CDMS catalog \citep{2005JMoSt.742..215M}; the entry was based on Fourier transform microwave (FTMW) spectroscopy data and \textit{ab initio} calculations reported by \citet{1998JChPh.109.3108B}.

In addition to the molecular structure, \citet{1998JChPh.109.3108B} also provide estimates of the electric dipole moment. However, the authors did not report the dipole polarizability, which is required for estimating reaction rate coefficients, as will be discussed in Section \ref{sec:discussion}. To this end, we carried out new calculations with the CFOUR (Coupled-Cluster techniques for Computational Chemistry) suite of electronic structure programs \citep{stanton_cfour_2017}, employing the coupled-cluster method with single, double, and perturbative triple excitations [CCSD(T)] under the frozen-core approximation, paired with a Dunning's correlation-consistent quadruple-$\zeta$ (cc-pVQZ) basis set. At this level of theory, we obtain an equilibrium dipole moment of $3.24\,\mathrm{D}$ in agreement with the value of $3.25\,\mathrm{D}$ obtained by \citet{1998JChPh.109.3108B} employing a smaller Dunning's triple-$\zeta$ (cc-pVTZ) basis set. The small change between the cc-pVTZ and cc-pVQZ values suggests that the one-electron properties have effectively converged with respect to basis set, thereby lending confidence in our calculations. With the same method and the cc-pVQZ basis set, we obtain a value of $10.3501\,\text{\AA}^3$ for the average dipole polarizability listed in Table~\ref{spec-par}.

\begin{deluxetable}{lccl}
    \tablewidth{\columnwidth}
    \tablecaption{Calculated dipole and polarizability for the related cyanopolyynes and isocyanopolyynes  \label{spec-par}}
    \tablehead{
        \colhead{Parameter} &\colhead{$\mu_e\ (\mathrm{D})$ \tablenotemark{a}}    &\colhead{$\alpha\ (\text{\AA}^3)$ \tablenotemark{b}}    &\colhead{Reference}}
    \startdata
        \ce{HC3N}   &3.788  &5.848    &\citet{2009ApJS..185..273W}\\
        \ce{HCCNC}   &2.990  &6.221    &\citet{2009ApJS..185..273W}\\
        \ce{HC5N}   &4.55  &9.61    &\citet{2014MNRAS.437..930L}\\
        \ce{HC4NC}   &3.24   &10.3501   & This work \\
        \ce{CH3C3N}   &5.041  &8.008    &\citet{2009ApJS..185..273W}
    \enddata
    \tablecomments{
    \tablenotetext{a}{The equilibrium electric dipole moment in units of Debye.}
    \tablenotetext{b}{The average dipole polarizability, in units of $\text{\AA}^3$.}
    }
\end{deluxetable}

\section{Observations \label{sec:results}}
The capabilities of the Green Bank Telescope (GBT) have expanded the molecular census in TMC-1 and, thereby, increased the known molecular inventory in the interstellar medium \citep{2017ApJ...843L..28M,2018Sci...359..202M}. The GBT observations of the GOTHAM project targeted the TMC-1 cyanopolyyne peak (CP) centered at $\alpha_\text{J2000}=\hms{04}{41}{42.5}$, $\delta_\text{J2000}=\dms{25}{41}{26.8}$, where the column densities of the carbon-chain species peak. The GOTHAM spectral line survey covers the GBT X-, K- and Ka-Bands with total 13.1 GHz frequency coverage from 7.906 to 29.827 GHz. The beam size varies between ${\sim}90 \arcsec$ at 8 GHz and ${\sim}26 \arcsec$ at 29 GHz. At a uniform $0.05 \mathrm{\ km\,s^{-1}}$ velocity resolution, the RMS noise ranges from ${\sim}2-20 \mathrm{\ mK}$ across the dataset. Detailed information concerning the GOTHAM observations and the data calibration can be found in \citet{McGuire:2020bb}. 

\begin{figure*}[!t]
    \centering
    \includegraphics[width=\textwidth]{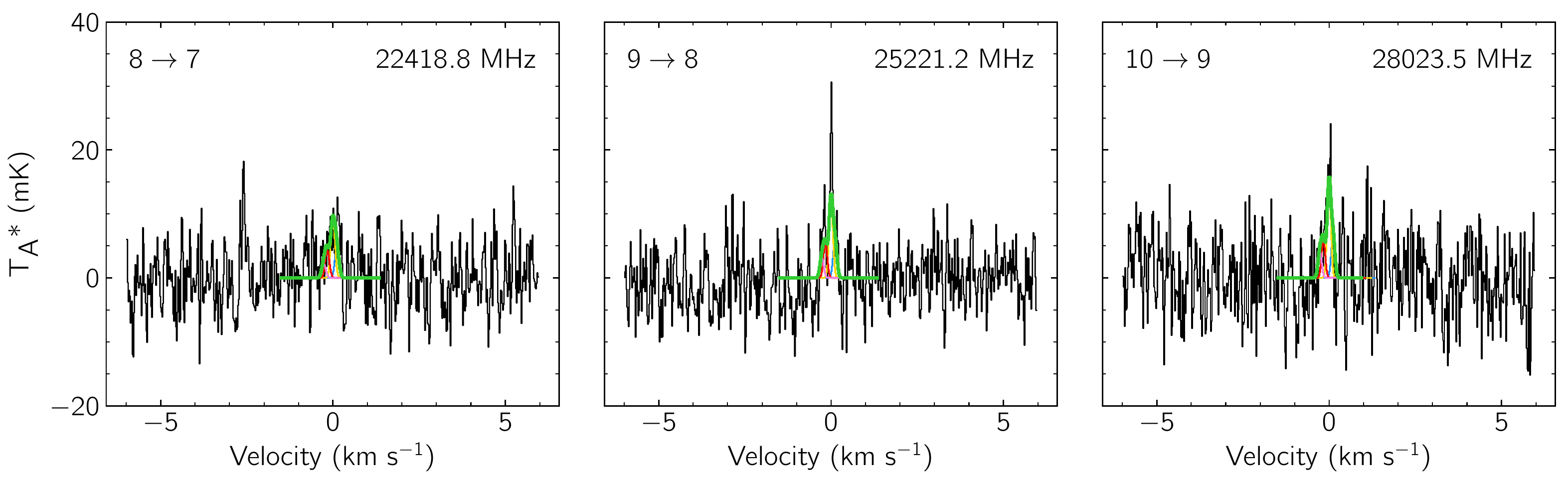}
    \includegraphics[width=\textwidth]{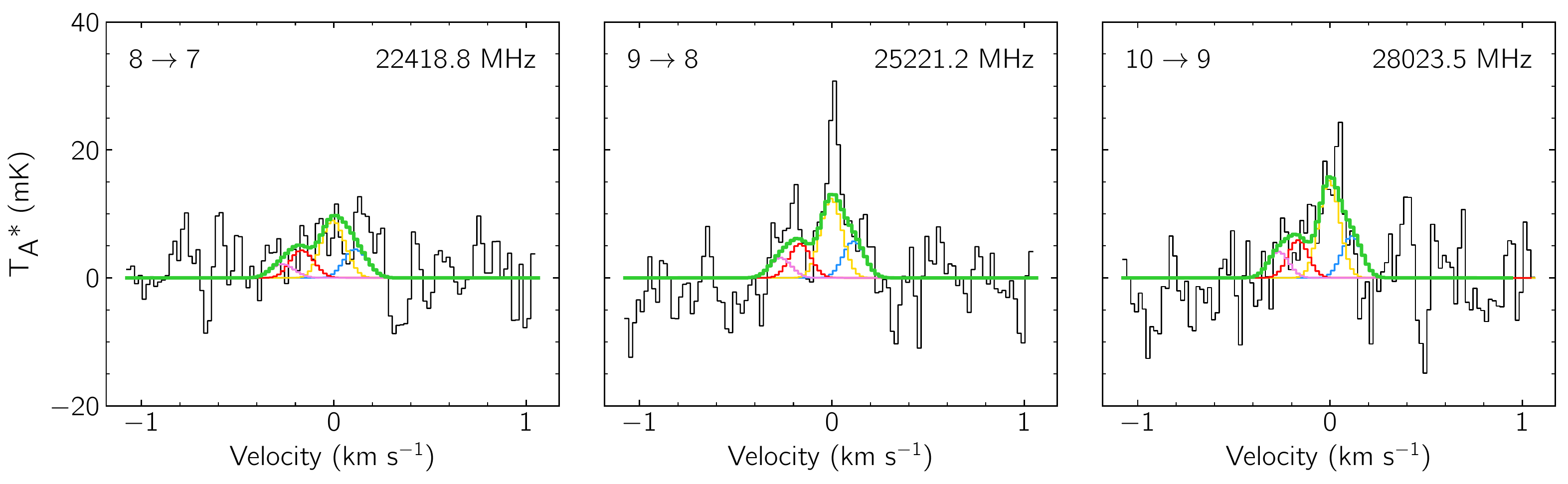}
    \caption{Individual line detections of \ce{HC4NC} in the GOTHAM data. The top row shows a wider view to provide context on the noise levels {, ${\sim}$5 mK}. The bottom row shows the same transitions, zoomed in to show detail. The spectra (black) are displayed in velocity space relative to 5.8\,km\,s$^{-1}$, and using the rest frequency given in the top right of each panel. Quantum numbers are given in the top left of each panel, neglecting hyperfine splitting. The best-fit model to the data, including all velocity components, is overlaid in green. Simulated spectra of the individual velocity components are shown in: blue (5.63\,km\,s$^{-1}$), yellow (5.75\,km\,s$^{-1}$), red (5.91\,km\,s$^{-1}$), and violet (6.01\,km\,s$^{-1}$). See Table~\ref{hc4nc_results}.}
    \label{spectra}
\end{figure*}

As presented in Figure~\ref{spectra}, we identified three emission features above the noise level of the observations assigned to \ce{HC4NC} with the GOTHAM survey. Each feature comprises three hyperfine components of the rotational transition. Table~\ref{tab:spc-par} summarizes the spectroscopic properties of the nine transitions. The \ce{HC4NC} lines show a good match between the observed frequencies and the calculated frequencies from the CDMS database assuming a systematic Local Standard of Rest velocity ($V_\mathrm{lsr}$) of $5.8\ \mathrm{km\ s^{-1}}$.

\begin{deluxetable}{cccccr}
    \tablecaption{Spectroscopic Properties of the Identified \ce{HC4NC} Lines \label{tab:spc-par}}
    \tablewidth{\columnwidth}
    \tablehead{
    \multicolumn{2}{c}{Transitions} &\colhead{Frequency}   &\colhead{$E_{up}$}   &\colhead{$\log_{10}{\frac{A_{ul}}{\mathrm{s^{-1}}}}$}   &\colhead{$S_{ij}\mu^{2}$}\\
    \colhead{$J'\ \rightarrow\ J''$}&\colhead{$F'\ \rightarrow\ F''$}&\colhead{(MHz)}&\colhead{(K)}&&\colhead{(D$^2$)}\\}
    \startdata
    $8\ \rightarrow\ 7$ &$9\ \rightarrow\ 8$   &22418.8438(10)   &4.84   &-6.1859   &94.43   \\
     &$8\ \rightarrow\ 7$   &22418.8461(10)   &4.84   &-6.1927   &83.18   \\
     &$7\ \rightarrow\ 6$   &22418.8498(10)   &4.84   &-6.1936   &73.24   \\
    $9\ \rightarrow\ 8$ &$10\ \rightarrow\ 9$   &25221.1790(16)   &6.05   &-6.0295   &105.07  \\
     &$9\ \rightarrow\ 8$   &25221.1808(17)   &6.05   &-6.0350   &93.88   \\
     &$8\ \rightarrow\ 7$   &25221.1837(17)   &6.05   &-6.0356   &83.88   \\
    $10\ \rightarrow\ 9$ &$11\ \rightarrow\ 10$   &28023.5067(25)   &7.40   &-5.8899   &115.69  \\
     &$10\ \rightarrow\ 9$   &28023.5082(26)   &7.40   &-5.8943   &104.57  \\
     &$9\ \rightarrow\ 8$   &28023.5105(26)   &7.40   &-5.8948   &94.52
    \enddata
    \tablecomments{The spectroscopic data of the \ce{HC4NC} transitions corresponding to the three detected emission features are taken from the CDMS catalogue \citep{2005JMoSt.742..215M} and the SPLATALOGUE spectroscopy database \footnote{\footnotesize{https://www.splatalogue.online}}.}
\end{deluxetable}

\subsection{Determinations of Column Density and Excitation Conditions \label{subsec:columndensity}}
%The three emission features assigned to \ce{HC4NC} were used to rigorously determine the molecular abundance and excitation conditions using the Markov chain Monte Carlo (MCMC) fitting method described in \citet{Loomis:2020aa}. Each of the spectral profiles consists of four individual velocity components \citep{Loomis:2020aa},
{A total of 13 transitions of \ce{HC4NC} (See Appendix \ref{apx:HC4NC})} were used to rigorously determine the molecular abundance and excitation conditions using the Markov chain Monte Carlo (MCMC) fitting method described in \citet{Loomis:2020aa}. Each of the {identified emission features} consists of four individual velocity components \citep{Loomis:2020aa}, indicating that TMC-1 is not quiescent and isotropic in terms of physical structure. This is supported by recent \ce{CCS} and \ce{HC3N} observations performed with the 45 m telescope at the Nobeyama Radio Observatory \citep{2018ApJ...864...82D}.

A uniform excitation temperature ($T_\mathrm{ex}$) and line width ($\Delta V$) for each velocity component are assumed, while source velocity ($V_\mathrm{lsr}$), source size, and column density ($N_\mathrm{T}$) are variable among different velocity components. Therefore, there are 14 free parameters in total to be adjusted in the MCMC analysis. A forward model with 14 free parameters is used to iteratively generate model spectra which are compared with the observations. Posterior probability distributions for each parameter and their covariances are generated via several million of these parameter draws, populating the corner plot in Appendix \ref{apx:HC4NC}. The resulting best-fit parameters of each velocity component of \ce{HC4NC} are summarized in Table~\ref{tab:obs-par}. {As shown in Figure~\ref{spectra}, if we take the noise level measured in each passband into account, the constructed profiles fit reasonably well with the observed spectra for the individual emission features.} A total $N_\mathrm{T}$ of $3.29^{+8.60}_{-1.20}\times 10^{11}\ \mathrm{cm^{-2}}$ with a $T_\mathrm{ex}$ of $6.7^{+0.3}_{-0.3} \mathrm{\ K}$ is determined for \ce{HC4NC}.

The $N_\mathrm{T}$ per velocity components show variation on the order of a factor of a few but have consistency in the order of magnitude, unlike the case of \ce{HC3N} and \ce{CCS} presented in \citet{2018ApJ...864...82D}. The variation of $N_\mathrm{T}$ arises from the degeneracy between the $N_\mathrm{T}$ of each component and its source size, found by the MCMC analysis. Without any spatial information to constrain the source sizes, we cannot conclude much about their chemical properties.

In addition to the \ce{HC4NC} analysis, we have also analyzed \ce{HCCNC} and \ce{HC6NC} in these observations; the results of these analyses are presented in Appendices \ref{apx:HCCNC} and \ref{apx:HC6NC}. \ce{HCCNC} is definitively detected with {six} emission features whereas there is no obvious emission detected for \ce{HC6NC}. The $N_\mathrm{T}$ for \ce{HCCNC} is measured to be $3.82^{+1.06}_{-0.53}\times 10^{12}\ \mathrm{cm^{-2}}$, while a $2\ \sigma$ upper limit of $<4.04\times 10^{11}\ \mathrm{cm^{-2}}$ for the \ce{HC6NC} column density is determined. \ce{HCCNC} has been previously detected in TMC-1 with Nobeyama 45-m observations \citep{2016ApJS..225...25G}, which reported $N_\mathrm{T}(\ce{HCCNC})$ to be $8.51^{+8.87}_{-1.9}\times 10^{12}\mathrm{cm^{-2}}$, consistent with our GOTHAM result. The column densities listed in Table~\ref{tab:nt-ratio} are the sums of the four detected velocity components, where the column densities of \ce{HC3N} and \ce{HC5N} are from \citet{Loomis:2020aa}. The detection of \ce{HCCNC} and \ce{HC4NC} in GOTHAM data gives column density ratios to their corresponding cyanide isomers of $2.2^{+0.7}_{-0.4}\%$ for \ce{HCCNC}$/$\ce{HC3N} and $0.49^{+1.32}_{-0.19}\%$ \ce{HC4NC}$/$\ce{HC5N} toward TMC-1. The observed results are used to constrain the reaction rate coefficients and branching ratios of the formation routes of \ce{HC4NC}, as will be discussed in Section \ref{subsec:model}.

\begin{table*}
\centering
\caption{\ce{HC4NC} best-fit parameters from the MCMC analysis \label{tab:obs-par}}
\begin{tabular}{c c c c c c}
\toprule
\multirow{2}{*}{Component}&	$v_{lsr}$					&	Size					&	\multicolumn{1}{c}{$N_T^\dagger$}					&	$T_{ex}$							&	$\Delta V$		\\
			&	(km s$^{-1}$)				&	($^{\prime\prime}$)		&	\multicolumn{1}{c}{(10$^{11}$ cm$^{-2}$)}		&	(K)								&	(km s$^{-1}$)	\\
\midrule
\hspace{0.1em}\vspace{-0.5em}\\
C1	&	$5.628^{+0.045}_{-0.038}$	&	$42^{+9}_{-9}$	&	$0.30^{+0.19}_{-0.13}$	&	\multirow{6}{*}{$6.7^{+0.3}_{-0.3}$}	&	\multirow{6}{*}{$0.120^{+0.012}_{-0.010}$}\\
\hspace{0.1em}\vspace{-0.5em}\\
C2	&	$5.745^{+0.021}_{-0.015}$	&	$21^{+7}_{-8}$	&	$1.35^{+1.38}_{-0.50}$	&		&	\\
\hspace{0.1em}\vspace{-0.5em}\\
C3	&	$5.907^{+0.038}_{-0.046}$	&	$62^{+20}_{-20}$	&	$0.23^{+0.12}_{-0.12}$	&		&	\\
\hspace{0.1em}\vspace{-0.5em}\\
C4	&	$6.009^{+0.044}_{-0.032}$	&	$9^{+11}_{-6}$	&	$1.40^{+8.48}_{-1.07}$	&		&	\\
\hspace{0.1em}\vspace{-0.5em}\\
\midrule
$N_T$ (Total)$^{\dagger\dagger}$	&	 \multicolumn{5}{c}{$3.29^{+8.60}_{-1.20}\times 10^{11}$~cm$^{-2}$}\\
\bottomrule
\end{tabular}

\begin{minipage}{0.75\textwidth}
	\footnotesize
	\textbf{Note} -- The quoted uncertainties represent the 16$^{th}$ and 84$^{th}$ percentile ($1\sigma$ for a Gaussian distribution) uncertainties.\\
	$^\dagger$Column density values are highly covariant with the derived source sizes. The marginalized uncertainties on the column densities are therefore dominated by the largely unconstrained nature of the source sizes, and not by the signal-to-noise of the observations. See Fig.~\ref{hc4nc_triangle} for a covariance plot, and \citet{Loomis:2020aa} for a detailed explanation of the methods used to constrain these quantities and derive the uncertainties.\\
	$^{\dagger\dagger}$Uncertainties derived by adding the uncertainties of the individual components in quadrature.
\end{minipage}
\label{hc4nc_results}
\end{table*}

\begin{deluxetable*}{llcccc}
    \tablecaption{Column Densities and XNC/XCN Ratios \label{tab:nt-ratio}}
    \tablewidth{0pt}
    \tablehead{
        \colhead{Species}   &\colhead{$N_\mathrm{T}$}   &\colhead{$N_\mathrm{T}$ with the Nobeyama Observations\tablenotemark{b}}  &\multicolumn{3}{c}{$N_\mathrm{T}(\ce{XNC})/N_\mathrm{T}(\ce{XCN})$}\\
                            &\colhead{($\mathrm{cm^{-2}}$)} &\colhead{($\mathrm{cm^{-2}}$)}    &\colhead{Observation}  &\colhead{High \ce{HC4NC} BF\tablenotemark{c}}    &\colhead{Low \ce{HC4NC} BF\tablenotemark{c}}
        }
    \startdata
    \ce{HCCCN}   &$1.75^{+0.05}_{-0.05}\times 10^{14}$\tablenotemark{a}   &$2.34^{+0.82}_{-0.30}\times 10^{14}$   &   &   &\\       
    \ce{HCCNC}  &$3.82^{+1.06}_{-0.53}\times 10^{12}$                    &$8.51^{+8.87}_{-1.90}\times 10^{12}$   &$2.2^{+0.7}_{-0.4}\%$&$3.0\%$&$3.0\%$\\
    \ce{HC4CN}   &$6.69^{+0.13}_{-0.13}\times 10^{13}$\tablenotemark{a}   &$5.89^{+1.52}_{-1.10}\times 10^{13}$   &   &   &\\
    \ce{HC4NC}  &$3.29^{+8.60}_{-1.20}\times 10^{11}$   &                                       &$0.49^{+1.32}_{-0.19}\%$&$2.6\%$&$0.34\%$\\
    \ce{HC6CN}   &$3.65^{+0.13}_{-0.12}\times 10^{13}$\tablenotemark{a}   &$4.57^{+1.74}_{-0.94}\times 10^{13}$   &   &   &\\
    \ce{HC6NC}  &$<4.04\times 10^{11}$   &   &$<1.1\%$    &   &
    \enddata
    \tablecomments{
    \tablenotetext{a}{\citet{Loomis:2020aa} estimated the column densities of cyanopolyynes with similar MCMC analyses of the GOTHAM data, assuming the four velocity components are cospatial.}
    \tablenotetext{b}{The column density estimated by the Bayesian approach of the spectral survey performed with the Nobeyama 45-m dish \citep{2016ApJS..225...25G}}
    \tablenotetext{c}{``High \ce{HC4NC} BF" corresponds to the model with a high branching fraction to form \ce{HC4NC} in the \ce{HC5NH+} dissociative recombination, i.e. shown in solid lines in Figure~\ref{fig:nautilus}, while ``Low \ce{HC4NC} BF" is the modeled result with a low branching fraction shown in dashed lines in Figure~\ref{fig:nautilus}.}
    }
\end{deluxetable*}

\subsection{Visualization of the Detection \label{subsec:vis}}
\begin{figure*}
    \centering
    \includegraphics[width=0.49\textwidth]{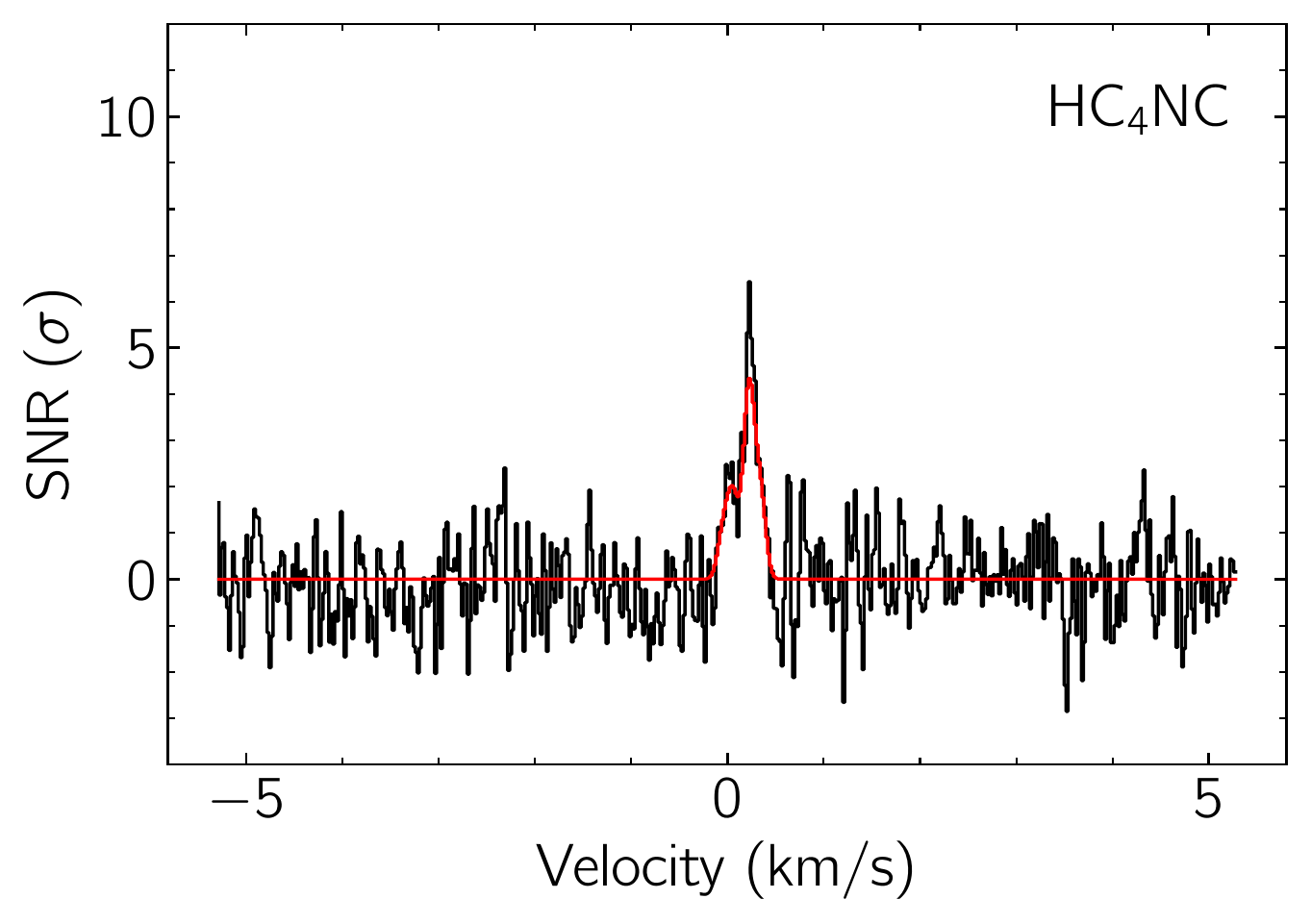}
    \includegraphics[width=0.49\textwidth]{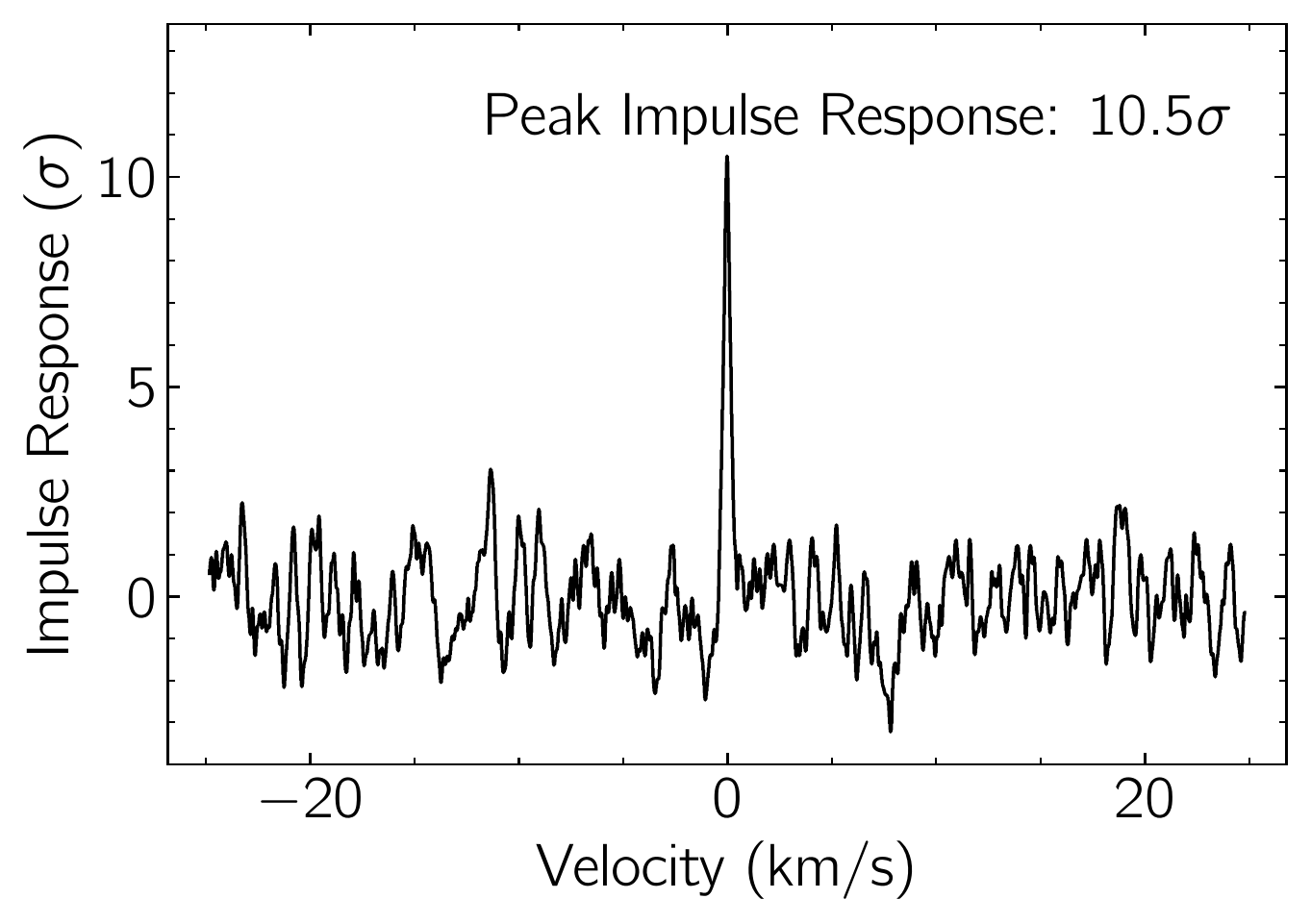}
    \caption{\emph{Left:} Velocity-stacked spectra of \ce{HC4NC} in black, with the corresponding stack of the simulation using the best-fit parameters to the individual lines in red. The data have been uniformly sampled to a resolution of 0.02\,km\,s$^{-1}$. The intensity scale is the signal-to-noise ratio of the spectrum at any given velocity. \emph{Right:} Impulse response function of the stacked spectrum using the simulated line profile as a matched filter. The intensity scale is the signal-to-noise ratio of the response function when centered at a given velocity. The peak of the impulse response function provides a minimum significance for the detection of 10.5$\sigma$. See \citet{Loomis:2020aa} for details.}
    \label{stack}
\end{figure*}

To better visualize the detection, and determine a minimum statistical significance, we constructed an intensity- and noise-weighted stacked composite spectrum using the GOTHAM data \citep{Loomis:2020aa}. The spectral stacking was performed {in velocity space} using {the 13} \ce{HC4NC} transitions covered by the survey. Another composite line profile using the best-fit parameters was constructed, and used as a matched filter to perform the cross-correlation and determine the statistical significance of the detection \citep{Loomis:2020aa}. The results are shown in Figure~\ref{stack}, and indicate a minimum significance to the detection of \ce{HC4NC} of $10.5\sigma$.

\section{Discussion \label{sec:discussion}}
\subsection{Chemical Networks}
A number of prior investigations have attempted to address the chemical origins of many of the cyanopolyynes observed in TMC-1 \citep{1998A&A...329.1156T, 2016ApJ...817..147T, 2018MNRAS.474.5068B}. For example, due to the significant abundance enhancement of \ce{HCC^{13}CN} relative to \ce{HC^{13}CCN} and \ce{H^{13}CCCN}, the formation of \ce{HC3N} was suggested to be dominated by the neutral-neutral reaction of \ce{C2H2} and the \ce{CN} radical \citep{1998A&A...329.1156T}. On the other hand, \ce{HC5N} and \ce{HC7N} show no such enhancement for the analogous \ce{^{13}C} position, implying that the primary formation route for \ce{HC5N} is the dissociative recombination (DR) reaction between the N-bearing hydrocarbon ions and electrons in cold environments \citep{2018MNRAS.474.5068B}. Furthermore, \citet{2014MNRAS.443..398L} pointed out that the \ce{H2CCN} + \ce{C} $\rightarrow$ \ce{HC3N} + \ce{H} reaction is also involved in producing \ce{HC3N}.

On the other hand, the chemistry of the corresponding isocyanopolyynes (\ce{HC_{2n}NC}) is less well known. Compared with neutral-neutral reactions, reaction schemes involving the DR process of protonated molecular ions such as \ce{HC3NH+} and \ce{HC2NCH+} are more likely to be the main production mechanisms for \ce{HCCNC} \citep{1992ApJ...386L..51K, 1997ApJ...479L..75G, 1999ApJ...519..697O, 2018MNRAS.474L..76V}. These protonated ions can be formed through ion-molecule reactions such as \ce{HCCH+ + HNC/HNC} and \ce{CH3CN + C+} \citep{1999ApJ...525..791T,2017MNRAS.470.3194Q}. Even though the chemistry of \ce{HC4NC} is less well-studied compared to \ce{HCCNC}, both of them belong to the same homologous series. We therefore assumed analogous formation schemes of \ce{HCCNC} and \ce{HC4NC}. In other words, \ce{HC4NC} would mainly form through the DR of the protonated cyanopolyynes \ce{HC5NH+} and protonated isocyanopolyynes \ce{HC4NCH+}.

One of the most prevalent destruction mechanisms of cyano- and isocyanopolyynes is ion-molecule chemistry, particularly reactions with \ce{C+}, \ce{H3+}, and \ce{HCO+} \citep{2009ApJS..185..273W}. In addition, as described in \citet{2014MNRAS.437..930L}, reactions with carbon atoms are also efficient. Therefore, we extrapolate the mechanisms involving carbon atoms to isocyanopolyynes and propose that the main destruction mechanisms of \ce{HC4NC} are with the ions mentioned above, neutral carbon, and photons.

In this work, we adopted the chemical network of kida.uva.2014 \citep{2015ApJS..217...20W}, modified as described in \citet{2018Sci...359..202M}, as the basis and added or updated the reactions related to \ce{HC3N}, \ce{HCCNC}, \ce{HC5N}, and \ce{HC4NC}. Note that we introduce \ce{HC4NC} as the only isomer of \ce{HC5N} in the network. We neglected the other \ce{HC5N} isomers to avoid adding more new species of which we have even less knowledge. In the following sections, we will discuss the choices and estimations of the reaction rate coefficients of the formation and destruction pathways of the four molecules of interest. The production and destruction routes regarding \ce{HC4NC} are summarized in Table~\ref{tab:reactions} with the corresponding rate coefficients.

\subsubsection{Formation Mechanisms -- The Dissociative Recombination Reactions}
The estimation of the branching ratios and rate coefficients of the \ce{HC3NH+} DR are constrained by the laboratory measurements of the DR of \ce{DC3ND+} and the consideration of isomerization among the products \citep[and references therein]{2019AnA...625A..91V}. Here, we adopted their values in this paper.

Since \ce{HC3N} and \ce{HCCNC} are both products of the \ce{HC3NH+} DR reactions, the \ce{HC5NH+} DR, originally included in the kida.uva.2014 network, is amended to include \ce{HC4NC} as another product:
\begin{align*}
        \ce{HC5NH+  + {$e$}-&-> C4H + HCN}      &22\%\\
                            &\ce{-> C4H + HNC}  &22\%\\
                            &\ce{-> CCH + HC3N} &6\%\\
                            &\ce{-> C5N + H2}   &4\%\\
                            &\ce{-> HC5N + H}   &43.8\%\\
                            &\ce{-> HC4NC + H}  &2.2\%
\end{align*}
The particular choice of the branching ratio is explained below.

Based on the potential energy surface of the various \ce{HC3N} isomers, \citet{2019AnA...625A..91V} suggested the branching fraction for the \ce{HC3NH+} DR forming \ce{HC3N} to be 20 times greater than that for the process forming \ce{HCCNC} (private communication with J. Loison). The energy difference between \ce{HC4NC} and \ce{HC5N} is calculated to be ${\sim}114.5$ kJ/mol (or 13771 K) with the W1BD thermochemical method, which is similar to the difference between \ce{HCCNC} and \ce{HC3N}, ${\sim}113.1$ kJ/mol (or 13603 K). Because of the lack of a laboratory measurement of the branching ratio in the \ce{HC5NH+} DR, we assume a fiducial ratio between the branching fractions for the \ce{HC5N} isomers to be 20, analogous to that of the \ce{HC3N} isomers. The total rate coefficient for the \ce{HC5NH+} DR, $2.0 \times 10^{-6} (T/300)^{-0.7} \mathrm{cm^3\ s^{-1}}$, and the branching ratios for the other product species are followed as suggested in kida.uva.2014.

The dissociative recombination of \ce{HC2NCH+} is another important pathway leading to \ce{HCCNC} \citep{FT9938902219}. In kida.uva.2014, the rate coefficient for the DR of \ce{HC2NCH+} is $6.0 \times 10^{-7} (T/300)^{-0.5} \mathrm{cm^3\ s^{-1}}$, which seems to be underestimated compared with the experimentally measured rate coefficient for the DR of \ce{DC3ND+}, $1.5 \times 10^{-6} (T/300)^{-0.7} \mathrm{cm^3\ s^{-1}}$ \citep{2004ApJ...613.1302G, 2012P&SS...60..102V}. We expect these rate coefficients to be similar because rate coefficients for DR tend to increase with complexity \citep{2012RPPh...75f6901L}, and because the two cations are of similar complexity. Considering that, we modified the total rate coefficient for the \ce{HC2NCH+} DR to be analogous with that of \ce{DC3ND+}.

Furthermore, we also added \ce{HC4NCH+} as secondary precursor of \ce{HC4NC}. For the \ce{HC4NCH+} DR, we assume the total rate coefficient to be consistent with the \ce{HC5NH+} DR rate coefficient of $2.0 \times 10^{-6} (T/300)^{-0.7} \mathrm{cm^3\ s^{-1}}$. The channels and branching ratios of the DR of \ce{HC4NCH+} are assumed to be equal to that of \ce{HC2NCH+} in the kida.uva.2014 network:
\begin{align*}
        \ce{HC4NCH+  + {$e$}-&-> C4H + HCN}     &38\%\\
                            &\ce{-> C4H2 + CN}  &38\%\\
                            &\ce{-> HC5N + H}   &4\%\\
                            &\ce{->HC4NC + H}   &20\%
\end{align*}
Note that, while \ce{HC4NC} can be protonated to form \ce{HC4NCH+}, the formation of \ce{HC4NCH+} is dominated by the proposed reaction between \ce{CH3C3N} and the \ce{C+} ion. Thus, consecutive protonation and de-protonation of \ce{HC4NC}, resulting in a zero net abundance change, is avoided.

Since the barrierless DR reactions contribute dominantly to the formation of isocyanopolyynes, we emphasize that the determination of the branching ratios are usually more crucial than those of the overall rate coefficients for the case of DR in astronomical environments \citep{2012RPPh...75f6901L}. Nonetheless, although the branching ratios of the related DR reactions are mostly estimated and relatively arbitrary due to the lack of experimental measurement other than for \ce{DC3ND+} \citep{2004ApJ...613.1302G}, we believe that the values we estimated are reasonable, as supported by the reproduction of observed values discussed below.

\subsubsection{Destruction Mechanisms}
As previously mentioned, the destruction of the cyano- and isocyanopolyynes is dominated by ion-molecule reactions and reactions with atomic carbon. The reaction coefficient of the related ion-molecule reactions are estimated with equation (3) from \citet{2009ApJS..185..273W}, which can be rewritten as
\begin{equation}
    k_\mathrm{D} = 0.4767 \frac{2 \pi e \mu_\mathrm{D}}{\sqrt{2 k T \mu}}
        + 0.62 \times 2 \pi e \sqrt{\frac{\alpha}{\mu}},
    \label{eqn:k_D}
\end{equation}
where $\mu_\mathrm{D}$ and $\alpha$ are the dipole moment and the average dipole polarizability of the neutral molecule, respectively, and $\mu$ is the reduced mass of the reactants. In addition to adding the new destruction routes proposed for \ce{HC4NC}, we also updated the ion-molecule reaction rate coefficients of \ce{HCCNC}, \ce{HC5N}, and \ce{CH3C3N} from the kida.uva.2014 network with this formula and the dipole moments and polarizabilities listed in Table~\ref{tab:reactions}. The reaction rate coefficients for the reactions of isocyanopolyynes with carbon atoms are estimated to be the same as those of the cyanopolyynes \citep{2014MNRAS.437..930L}, while the reaction coefficients for the UV photon dissociation and cosmic-ray ionization reactions of \ce{HC4NC} are assumed to be the same as those of \ce{HCCNC} in kida.uva.2014 respectively.
%photon dissociation reactions (Heays et al. 2017)

\subsection{Chemical Modeling \label{subsec:model}}
We used the three-phase gas-grain astrochemical model \texttt{NAUTILUS 1.1} \citep{2016MNRAS.459.3756R} together with our updated network to attempt to reproduce the abundances of \ce{HC4NC} and the related species. Physical conditions are assumed to follow typical cold dense cloud conditions, i.e. a gas and dust temperature of 10 K, a gas density $n_{\rm H}$ of $2\times10^4~\mathrm{cm}^{-3}$, a visual extinction ($A_{\rm v}$) of 10, and a cosmic ray ionization rate ($\zeta$) of $1.3\times10^{-17} \mathrm{s}^{-1}$ \citep{2016MNRAS.459.3756R}. We adopted assumed initial elemental abundances in TMC-1 CP as described in \citet{2011A&A...530A..61H} with the exception of atomic oxygen. The resulting abundances, with respect to the $N_\mathrm{T, (\ce{H2})} \sim 10^{22} \mathrm{cm^{-2}}$ \citep{2016ApJS..225...25G}, were converted to column densities and compared with the observed values.

\begin{figure}
    \centering
    \includegraphics[width=\columnwidth]{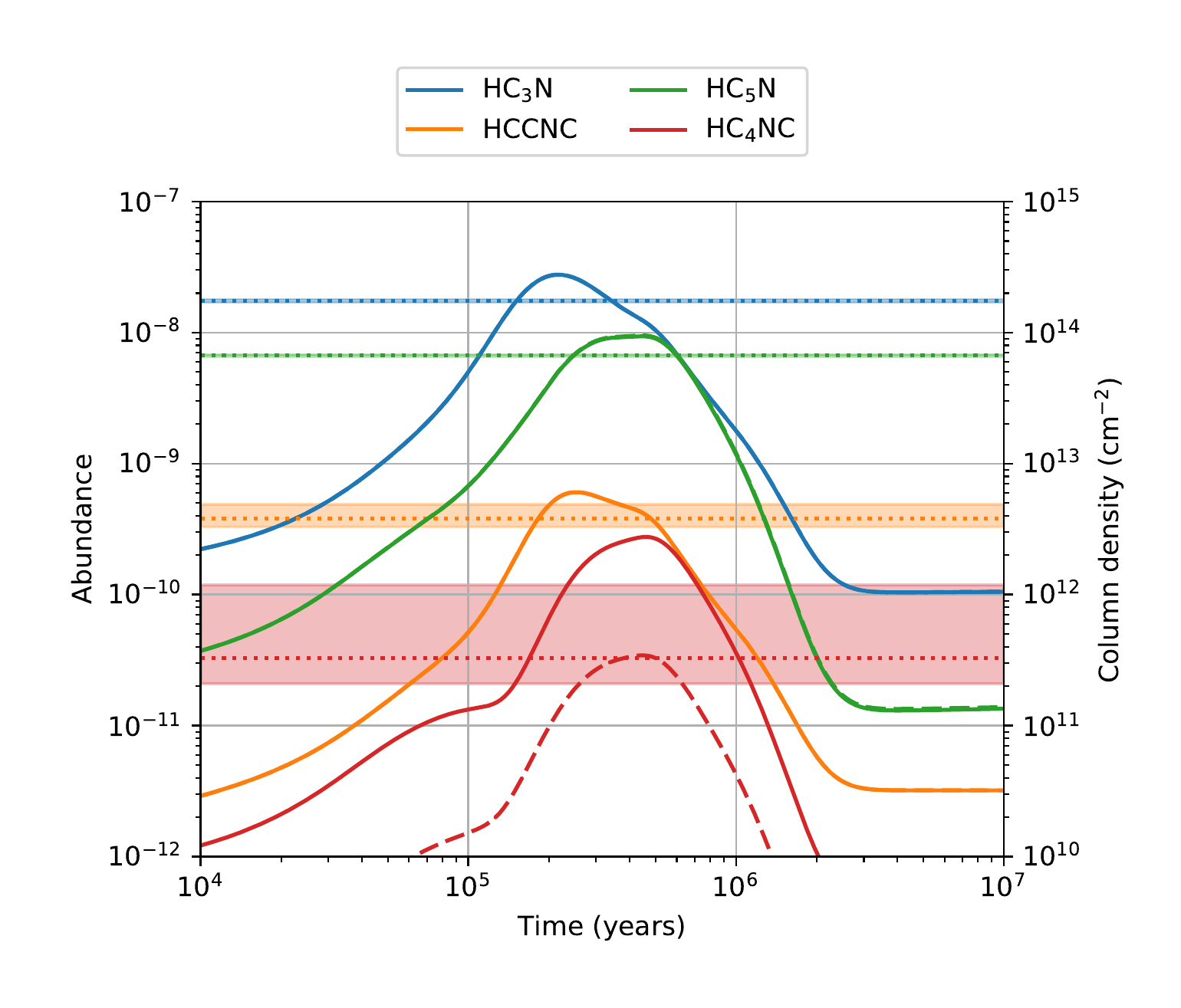}
    \caption{The observed and predicted abundances of \ce{HC3N}, \ce{HC5N}, \ce{HCCNC}, and \ce{HC4NC} are shown in blue, orange, green, and red respectively. The dotted lines and the shaded regions correspond to the mean and the $1 \sigma$ range of the observed abundances. The solid and dashed lines represent two models with \ce{HC5N}/\ce{HC4NC} branching ratios for \ce{HC5NH+} DR of 20 and 200 respectively.} Note that the predicted abundances of \ce{HC3N}, \ce{HC5N}, and \ce{HCCNC} from the two models overlap and are indistinguishable in the figures.
    \label{fig:nautilus}
\end{figure}

We found that both cyano- and isocyanopolyynes are highly sensitive to the initial oxygen elemental abundance. A higher oxygen abundance would result in lower abundances of the \ce{HC3N}, \ce{HCCNC}, \ce{HC5N}, and \ce{HC4NC} molecules because the majority of \ce{C} is being locked into \ce{CO} while reacting with the abundant \ce{O}. In Figure~\ref{fig:nautilus}, we present the results of the chemical modeling with an initial C/O ratio of 1.1, in which the model at an age of ${\sim}3.5\times 10^{5}$ yr gives satisfactory agreement with the observations for \ce{HC3N, HCCNC, and HC5N}. The initial physical conditions and elemental abundances are all homogeneous among the current series of GOTHAM papers \citep{McGuire:2020aa,Loomis:2020aa,McCarthy:2020aa,Xue:2020aa,Burkhardt:2020aa} and have reproduced the observed abundances of the other cyanopolyynes species \ce{HC7N}, \ce{HC9N}, and \ce{HC11N} well. Compared with previous astrochemical modelling on TMC-1, the modelled results produce a similar agreement. For example, in \citet{2014MNRAS.437..930L}, when assuming the C/O ratio to be 0.95, the peak abundances for \ce{HC3N} and \ce{HC5N} are ${\sim} 4\times10^{-8}$ and ${\sim}7\times10^{-9}$ respectively and occur at ${\sim}3\times10^{5}\ \mathrm{yr}$, which are consistent with our results.

The overproduction of \ce{HC4NC} could be explained by the defects in the chemical network. Concerning destruction, there could be secondary destruction mechanisms that we have not accounted for, while concerning production, the branching ratios in the related DR processes could be inaccurate. Firstly, the ratio between the branching fraction for forming \ce{HC5N} and that for forming \ce{HC4NC} in the \ce{HC5NH+} DR was assumed to be an analogous value of 20 from the \ce{HC3NH+} DR, which could be underestimated. We conducted additional models by varying this ratio and found that increasing it would result in a significant decrease in the simulated abundance of \ce{HC4NC} while the increase in \ce{HC5N} is less significant, as shown in Figure~\ref{fig:nautilus}. When this ratio is set to be 200, the modeled abundance ratio for \ce{HC4NC}$/$\ce{HC5N} can reach $0.34\%$, which matches well with the observed value, $0.49^{+1.32}_{-0.19}\%$. Therefore, as constrained by the observed abundances, this ratio is suggested to fall within a range of 20 to 200. Secondly, neglecting other possible \ce{HC5N} isomers in the DR processes would also lead to an overestimation of the branching fractions for forming \ce{HC4NC} and \ce{HC5N} in the \ce{HC4NCH+} DR. A reduction in the branching ratio could easily reduce the simulated \ce{HC4NC} abundance. Experimental studies on the formation and destruction pathways of this molecule are rare, and its detection in TMC-1 therefore highlights the need for more experimental and theoretical work

\subsection{CN/NC Formation Chemistry \label{subsec:nc-cn}}
In the current study, we have assumed the formation mechanism of \ce{HC4NC} to be analogous to that of \ce{HCCNC} with the understanding that \ce{HC3N} and \ce{HC5N}, and thus \ce{HCCNC} and \ce{HC4NC}, might have different dominant formation pathways. As such, the model results presented here are only a first attempt at understanding this chemistry with the knowledge that refinements to the models will be necessary as more experimental studies become available.

The current model shows that the HC$_{\rm n}$NH$^{+}$ DR is the dominant pathway in the formation of \ce{HCCNC} and \ce{HC4NC}, whereas there are several reaction channels contributing to the \ce{HCCCN} and \ce{HC4CN} production and different pathways dominate at different times, in disagreement with what the \ce{^{13}C}-isotopologue observation suggests. The resultant model abundance ratios are comparable for \ce{HCCNC}$/$\ce{HCCCN} ($\sim 3.0\%$) and \ce{HC4NC}$/$\ce{HC4CN} ($\sim 2.6\%$) at $\sim 3.5\times 10^{5}$ yr. 

In contrast, the observed \ce{HC4NC}$/$\ce{HC4CN} abundance ratio in TMC-1, $0.49^{+1.32}_{-0.19}\%$, is lower than the \ce{HCCNC}$/$\ce{HCCCN} abundance ratio, $2.2^{+0.7}_{-0.4}\%$, within 1$\sigma$ uncertainty. The uncertainties in the observed ratios are largely introduced by the poor constraint on the spatial distribution of these molecules. One caveat is that, as \citet{2005ApJ...632..333R} highlighted, a necessary prerequisite to interpret the relative abundance ratio between any molecular species detected in astronomical environments, including cyanide and isocyanide isomers, is that they must be co-spatial.
 
A subsequent dedicated search for cyanide and isocyanide pairs in different interstellar sources is justified, because the abundance ratio between cyanide and isocyanide isomers could also vary among sources. For example, the \ce{HCCNC}$/$\ce{HCCCN} abundance ratio toward the L1544 pre-stellar core, $\sim(3.5 - 13.8)\%$, is elevated relative to that in the TMC-1 dark cloud \citep{2018MNRAS.474L..76V}. Compared with TMC-1, L1544 is at a later stage along the path of star formation and has a slightly higher excitation temperature of 6 - 8 K \citep{2018MNRAS.474L..76V}. Determining the cause of the variation in CN/NC isomeric ratios may prove useful in constraining the dominant pathways and their dependence on the physico-chemical history of the source.

In addition, studies on other cyanide/isocyanide isomers in TMC-1 would help to address how the \ce{NC}$/$\ce{CN} ratio varies among different pairs of species, such as \ce{CH3CN} and \ce{CH3NC}. To date, only \ce{CH3CN} has been detected towards TMC-1 \citep{1984ApJ...282..516I, 2016ApJS..225...25G}, while \ce{CH3NC} may be detected as the GOTHAM survey progresses.

\startlongtable
\begin{deluxetable*}{lllrcl}
    \tablewidth{0pt}
    \tablecaption{Summary of the Proposed Dominant Reactions For \ce{HC4NC} \label{tab:reactions}}
    \tablehead{
        \colhead{Reactions} & \colhead{$\alpha$} & \colhead{$\beta$}  & \colhead{$\gamma$} & \colhead{Formula Type} & \colhead{$k (10 \mathrm{K})$}
    }
    \startdata
        \sidehead{Production Routes:}
        \ce{HC5NH+  + {$e$}- -> HC4NC   + H}                  & $4.400\times10^{-8}$ & -0.7 & 0 & 3     &$4.758\times10^{-7}\ $\tablenotemark{a}\\
        \ce{HC4NCH+ + {$e$}- -> HC4NC   + H}\tablenotemark{b} & $4.000\times10^{-7}$ & -0.7 & 0 & 3     &$4.326\times10^{-6}\ $\tablenotemark{c}\\
        \sidehead{Destruction Routes:}
	    \ce{HC4NC   + C+     -> HC5N+   + C}                  & 0.2   & $2.334\times10^{-9}$ & 3.499 & 4 &$4.554\times10^{-9}\ $\tablenotemark{d}\\
	    \ce{HC4NC   + C+     -> CNC+    + C4H}                & 0.2   & $2.334\times10^{-9}$ & 3.499 & 4 &$4.554\times10^{-9}\ $\tablenotemark{d}\\
	    \ce{HC4NC   + C+     -> C6N+    + H}                  & 0.2   & $2.334\times10^{-9}$ & 3.499 & 4 &$4.554\times10^{-9}\ $\tablenotemark{d}\\
	    \ce{HC4NC   + C+     -> C5H+    + CN}                 & 0.2   & $2.334\times10^{-9}$ & 3.499 & 4 &$4.554\times10^{-9}\ $\tablenotemark{d}\\
	    \ce{HC4NC   + C+     -> C4H+    + CCN}                & 0.2   & $2.334\times10^{-9}$ & 3.499 & 4 &$4.554\times10^{-9}\ $\tablenotemark{d}\\
	    \ce{HC4NC   + H3+    -> HC4NCH+ + H2}                 & 1.0   & $4.420\times10^{-9}$ & 3.499 & 4 &$4.312\times10^{-8}\ $\tablenotemark{d}\\
	    \ce{HC4NC   + HCO+   -> HC4NCH+ + CO}                 & 1.0   & $1.642\times10^{-9}$ & 3.499 & 4 &$1.602\times10^{-8}\ $\tablenotemark{d}\\
	    \ce{HC4NC   + H3O+   -> HC4NCH+ + H2O}                & 1.0   & $1.928\times10^{-9}$ & 3.499 & 4 &$1.881\times10^{-8}\ $\tablenotemark{d}\\
	    \ce{HC4NC   + H+     -> CN      + C4H2+}              & 0.333 & $7.557\times10^{-9}$ & 3.499 & 4 &$2.455\times10^{-8}\ $\tablenotemark{d}\\
	    \ce{HC4NC   + H+     -> H2      + C5N+}               & 0.333 & $7.557\times10^{-9}$ & 3.499 & 4 &$2.455\times10^{-8}\ $\tablenotemark{d}\\
	    \ce{HC4NC   + H+     -> C       + H2C4N+}             & 0.333 & $7.557\times10^{-9}$ & 3.499 & 4 &$2.455\times10^{-8}\ $\tablenotemark{d}\\
	    \ce{HC4NC   + He+    -> He      + C4H   + CN+}        & 0.5   & $3.852\times10^{-9}$ & 3.499 & 4 &$1.879\times10^{-8}\ $\tablenotemark{d}\\
	    \ce{HC4NC   + He+    -> He      + C4H+  + CN}         & 0.5   & $3.852\times10^{-9}$ & 3.499 & 4 &$1.879\times10^{-8}\ $\tablenotemark{d}\\
	    \ce{HC4NC   + C      -> C      + HC5N}                & $1.000\times10^{-10}$   & 0 & 0 & 3       &$1.000\times10^{-10}\ $\tablenotemark{e}\\
	    \ce{HC4NC   + CRPh    -> CN     + C4H}                & $3.450\times10^{3}$   & 0 & 0 & 1       &$4.485\times10^{-14}\ $\tablenotemark{f}\\
	    \ce{HC4NC   + Photon -> CN     + C4H}                 & $9.540\times10^{-10}$ & 0 & 1.830 & 2   &$1.076\times10^{-17}\ $\tablenotemark{f}
    \enddata
    \tablecomments{Definitions of $\alpha$, $\beta$, and $\gamma$ can be found on the KIDA online database (\url{http://kida.astrophy.u-bordeaux.fr/help.html}). Formulae of type 1 and 2 are $k = \alpha \zeta$ and $k = \alpha e^{-\gamma A_\nu}$, where $k$ is in $\mathrm{s^{-1}}$, and formulae of type 3 and 4 are $k(T) = \alpha \left(T/300\right)^\beta e^{-\gamma/T}$ and $k(T) = \alpha \beta \left(0.62 + 0.4767 \gamma \left(300/T\right)^{0.5}\right)$, where $k$ is in $\mathrm{cm^3\,s^{-1}}$ and $T$ is in $\mathrm{K}$, respectively.
    \tablenotetext{a}{The total reaction coefficient of the \ce{HC5NH+} DR is $2.0 \times 10^{-6} (T/300)^{-0.7} \mathrm{cm^3\ s^{-1}}$ followed as suggested in kida.uva.2014 while the branching ratio leading to \ce{HC4NC} is assumed to be one-twentieth of that leading to \ce{HC5N}, which is based on the branching fractions for producing \ce{HC3N} and \ce{HCCNC} of the \ce{HC3NH+} DR \citep{2019AnA...625A..91V}.}
    \tablenotetext{b}{\ce{HC4NCH+} is mainly produced through the reaction between \ce{CH3C3N} and the \ce{C+} ion.}
    \tablenotetext{c}{The reaction coefficient of the \ce{HC4NCH+} DR is assumed to have the same rate coefficient as that of \ce{HC5NH+}: $2.0 \times 10^{-6} (T/300)^{-0.7} \mathrm{cm^3\ s^{-1}}$, while the branching ratio is assumed to be similar to that of \ce{HC2NCH+}, which is included in kida.uva.2014.}
    \tablenotetext{d}{Rate coefficient estimated from Equation (\ref{eqn:k_D}) with $\mu$ of $3.24\ \mathrm{D}$ and $\alpha$ of $10.3501\ \text{\AA}^3$.}
    \tablenotetext{e}{Rate coefficient same as the 
    family reacting with atomic carbon \citep{2014MNRAS.437..930L}.}
    \tablenotetext{f}{Rate coefficient same as that of the $\ce{HCCNC} + \mathrm{CRPh}$ and $\ce{HCCNC} + \mathrm{Photon}$ reactions in kida.uva.2014 respectively.}
    }
\end{deluxetable*}

\section{Summary \label{sec:conclusions}}
In this paper, we report the astronomical detection of \ce{HC4NC} for the first time in the interstellar medium using the GOTHAM survey at a minimum significance of 10.5$\sigma$. Three emission features above the noise level of the observations are assigned to \ce{HC4NC}. Our analysis indicates a total of four distinct velocity components contribute to the emission signal observed for this species. The observed ratio between \ce{HC4NC} and its cyanopolyyne counterpart \ce{HC5N} is $\sim 0.49^{+1.32}_{-0.19}\%$ while the observed relative abundance ratio between \ce{HCCNC} and \ce{HC3N} is $\sim 2.2^{+0.7}_{-0.4}\%$.

The synthesis of the \ce{HC4NC} molecule is linked to the chemistry of the protonated cyanides and isocyanides. We attempted to reproduce the observed abundances of the selected cyano- and isocyanopolyynes with the inclusion of dissociative recombination as major formation routes and ion-molecule reactions, as well as reactions with atomic carbon as dominant destruction routes. We are aware that \ce{HC3N} and \ce{HC5N} have different dominant formation pathways whereas the chemical network of \ce{HC4NC} in the current study is assumed to be analogous to that of \ce{HCCNC}. The similar molecular structure of the two isocyanopolyynes makes it the best assumption we can posit.

The chemical modelling presented reproduces the observed abundance of \ce{HC4NC} within an order of magnitude. The result of the chemical modelling suggests that the considered formation and destruction routes are reasonable and relevant for \ce{HC4NC} and has enabled us to constrain the reaction rate coefficients to some extent. With the increasing number of detected cyano- and isocyanopolyynes in astronomical environments, accurate laboratory measurements of the rate coefficients and branching ratios for reactions of interest would certainly help to better reproduce the observed results.

\acknowledgments

A.M.B. acknowledges support from the Smithsonian Institution as a Submillimeter Array (SMA) Fellow. M.C.M and K.L.K.L. acknowledge financial support from NSF grants AST-1908576, AST-1615847, and NASA grant 80NSSC18K0396. Support for B.A.M. was provided by NASA through Hubble Fellowship grant \#HST-HF2-51396 awarded by the Space Telescope Science Institute, which is operated by the Association of Universities for Research in Astronomy, Inc., for NASA, under contract NAS5-26555. C.N.S. thanks the Alexander von Humboldt Stiftung/Foundation for their generous support, as well as V. Wakelam for use of the \texttt{NAUTILUS} v1.1 code. C.X. is a Grote Reber Fellow, and support for this work was provided by the NSF through the Grote Reber Fellowship Program administered by Associated Universities, Inc./National Radio Astronomy Observatory and the Virginia Space Grant Consortium. E.H. thanks the National Science Foundation for support through grant AST 1906489. S.B.C. and M.A.C. were supported by the NASA Astrobiology Institute through the Goddard Center for Astrobiology. The National Radio Astronomy Observatory is a facility of the National Science Foundation operated under cooperative agreement by Associated Universities, Inc. The Green Bank Observatory is a facility of the National Science Foundation operated under cooperative agreement by Associated Universities, Inc.

\appendix

\section{MCMC Fitting Detail for \texorpdfstring{\ce{HC4NC}}{HC4NC}\label{apx:HC4NC}}
\begin{figure*}[hbt!]
\centering
\includegraphics[width=\textwidth]{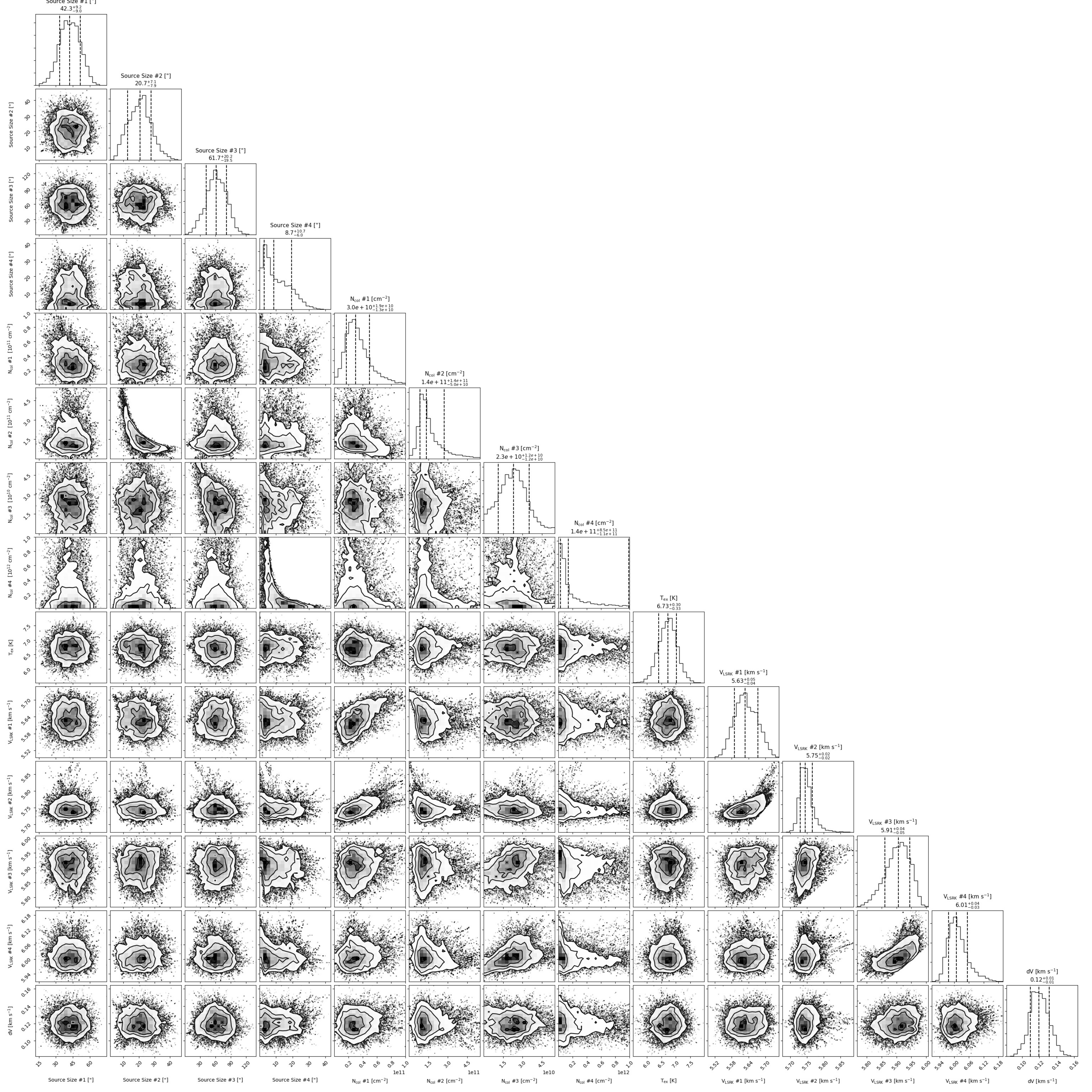}
\caption{Parameter covariances and marginalized posterior distributions for the \ce{HC4NC} MCMC fit. 16$^{th}$, 50$^{th}$, and 84$^{th}$ confidence intervals (corresponding to $\pm$1 sigma for a Gaussian posterior distribution) are shown as vertical lines. }
\label{hc4nc_triangle}
\end{figure*}

\restartappendixnumbering
\textbf{A total of 13 transitions (including hyperfine components) of \ce{HC4NC} were covered by GOTHAM observations at the time of analysis and were above the predicted flux threshold of 5\%, as discussed in \citet{Loomis:2020aa}. Of these transitions, none were coincident with interfering transitions of other species, and thus a total of 13 transitions were considered. Observational data windowed around these transitions, spectroscopic properties of each transition, and the partition function used in the MCMC analysis are provided in the Harvard Dataverse repository \citep{GOTHAMDR1}.} A corner plot of the parameter covariances and their distribution for the \ce{HC4NC} MCMC fit is shown in Figure~\ref{hc4nc_triangle}. Worth noting are the strong covariances between the column density and the source size for sources \#2 and \#4. The poor constraint on these source sizes leads to a large uncertainty in the total column density. Future detections of lines at lower or higher frequencies to anchor the source size fit (through measured beam dilution) would greatly enhance the precision of the column density measurement.

\section{\texorpdfstring{\ce{HCCNC}}{HCCNC} Analysis Results \label{apx:HCCNC}}
\begin{figure*}[!t]
    \centering
    \includegraphics[height=0.25\textheight]{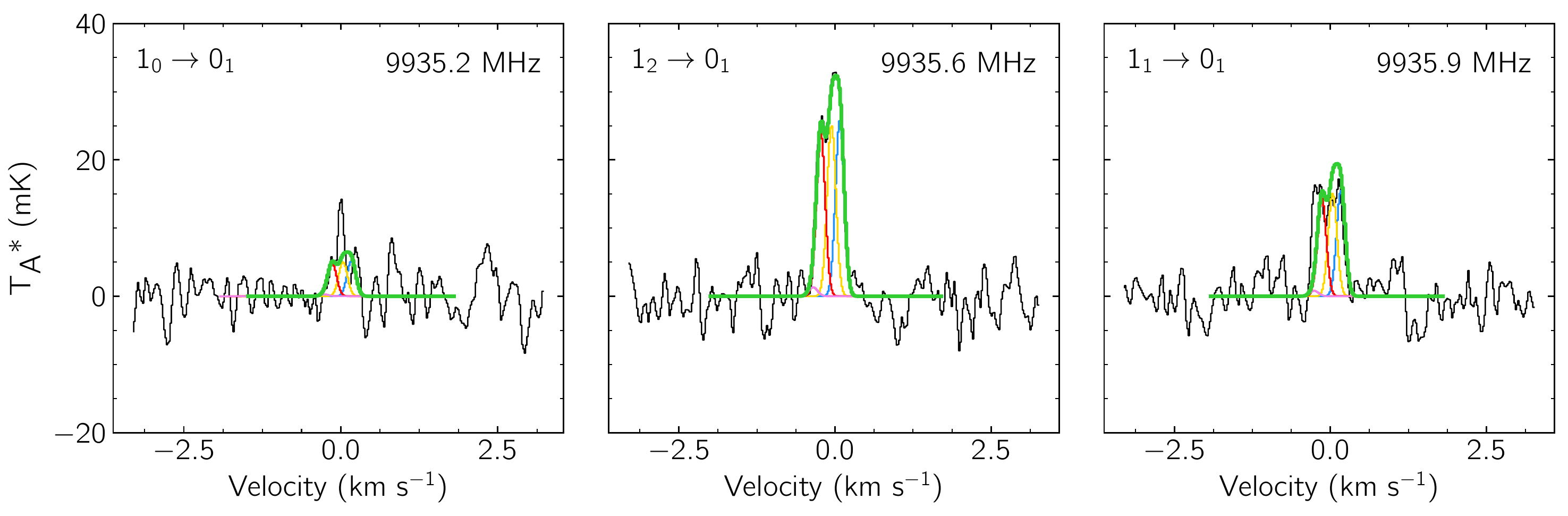}
    \includegraphics[height=0.25\textheight]{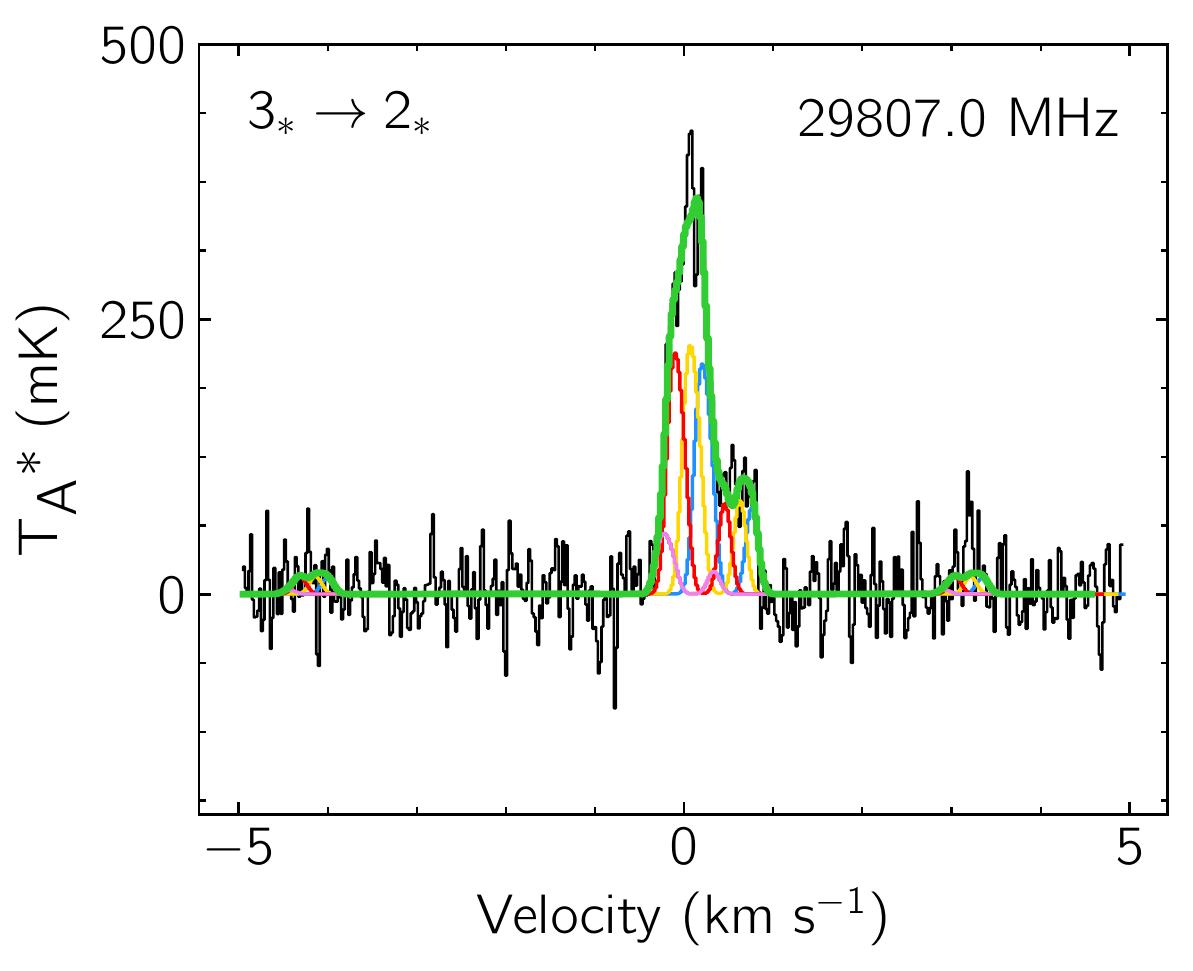}
    \caption{Similar with Figure~\ref{spectra}. Individual line detections of \ce{HCCNC} in the GOTHAM data. The observed spectra (black) are displayed in velocity space relative to 5.8\,km\,s$^{-1}$ and the simulated spectra of the individual velocity components are shown in blue (5.62\,km\,s$^{-1}$), yellow (5.76\,km\,s$^{-1}$), red (5.93\,km\,s$^{-1}$), and violet (6.05\,km\,s$^{-1}$), which are summarized in Table~\ref{hc2nc_results}, {with the best-fit model including all velocity components overlaid in green}.}
    \label{hc2nc_spectra}
\end{figure*}

\restartappendixnumbering
{An identical analysis to that for \ce{HC4NC} was carried out for \ce{HCCNC}. Six emission features contributed by the nine rotational transitions (including hyperfine components) of \ce{HCCNC} are well-detected above the noise, as shown in Figure~\ref{hc2nc_spectra}.} The top three panels are the three hyperfine components of the 1 -- 0 transition respectively while the bottom panel shows all the hyperfine components of the 3 -- 2 transition. The spectroscopic properties of the nine transitions are summarized in Table~\ref{tab:spc-par-hc2nc}.

\begin{figure*}
    \centering
    \includegraphics[width=0.49\textwidth]{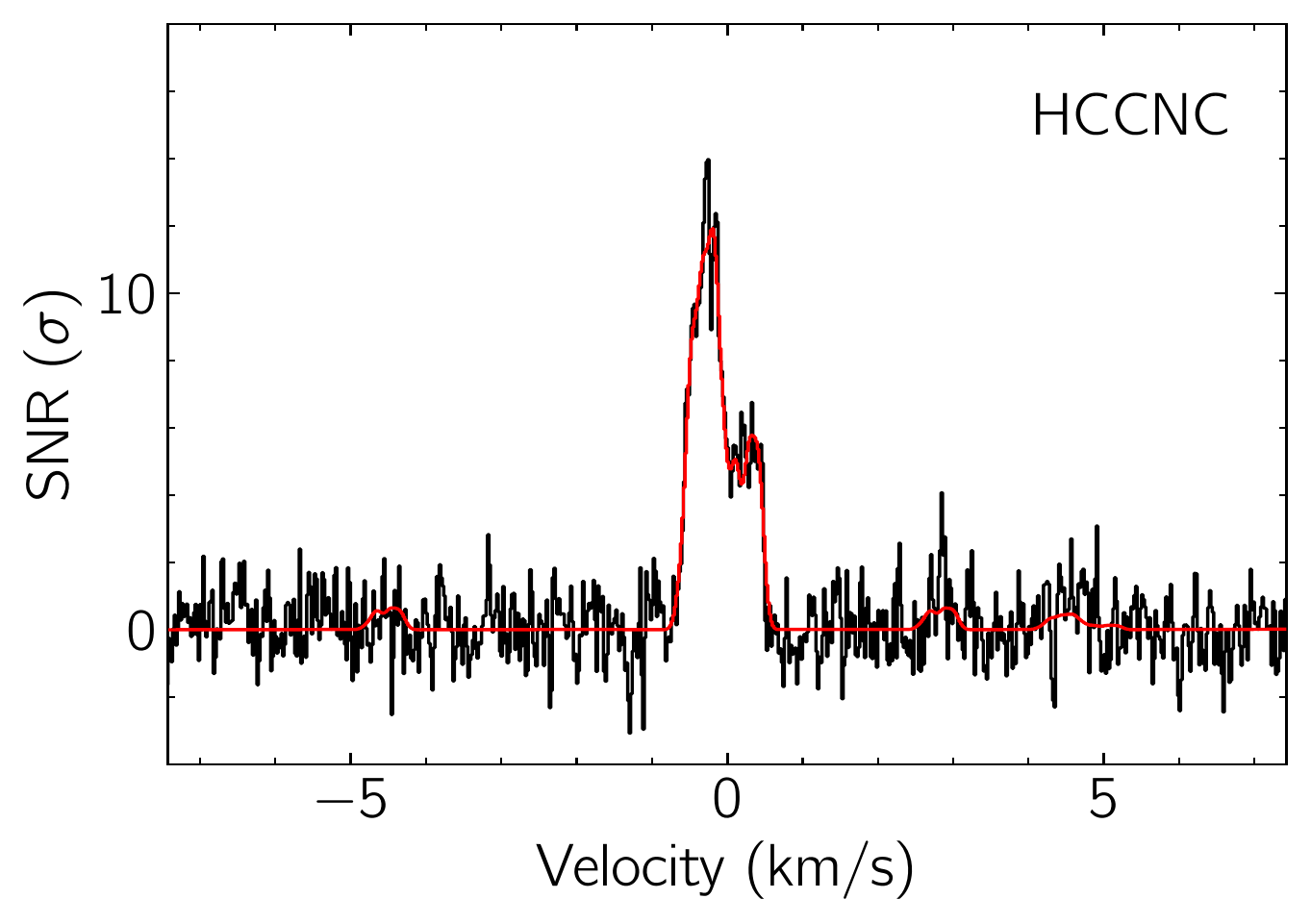}
    \includegraphics[width=0.49\textwidth]{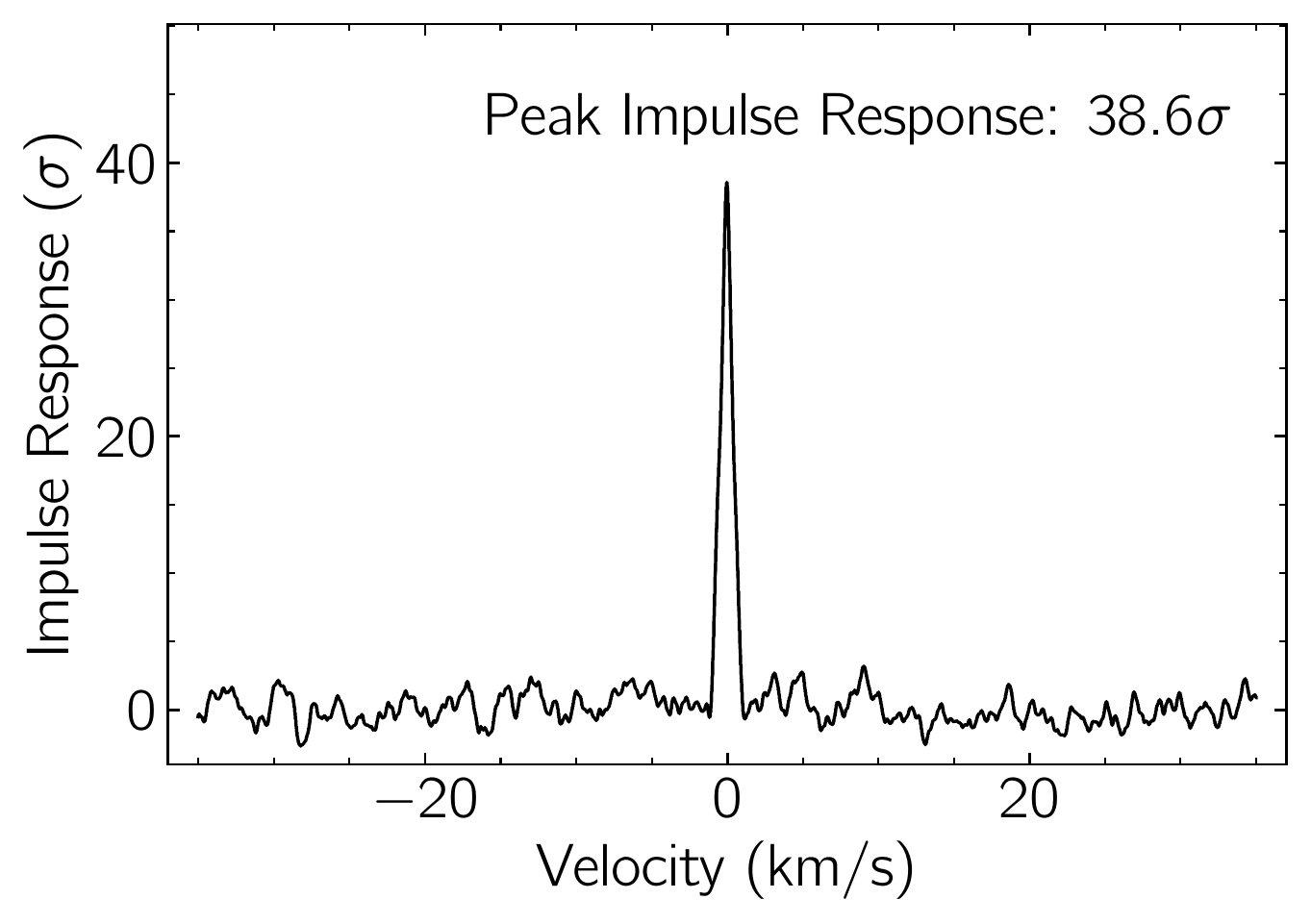}
    \caption{Similar with Figure~\ref{stack}. \emph{Left:} Velocity-stacked spectra of \ce{HCCNC} in black, with the corresponding stack of the simulation using the best-fit parameters to the individual lines in red. \emph{Right:} Impulse response function of the stacked spectrum using the simulated line profile as a matched filter. The peak of the impulse response function provides a minimum significance for the detection of 38.6$\sigma$.}
    \label{hc2nc_stack}
\end{figure*}

\begin{figure*}
\centering
\includegraphics[width=\textwidth]{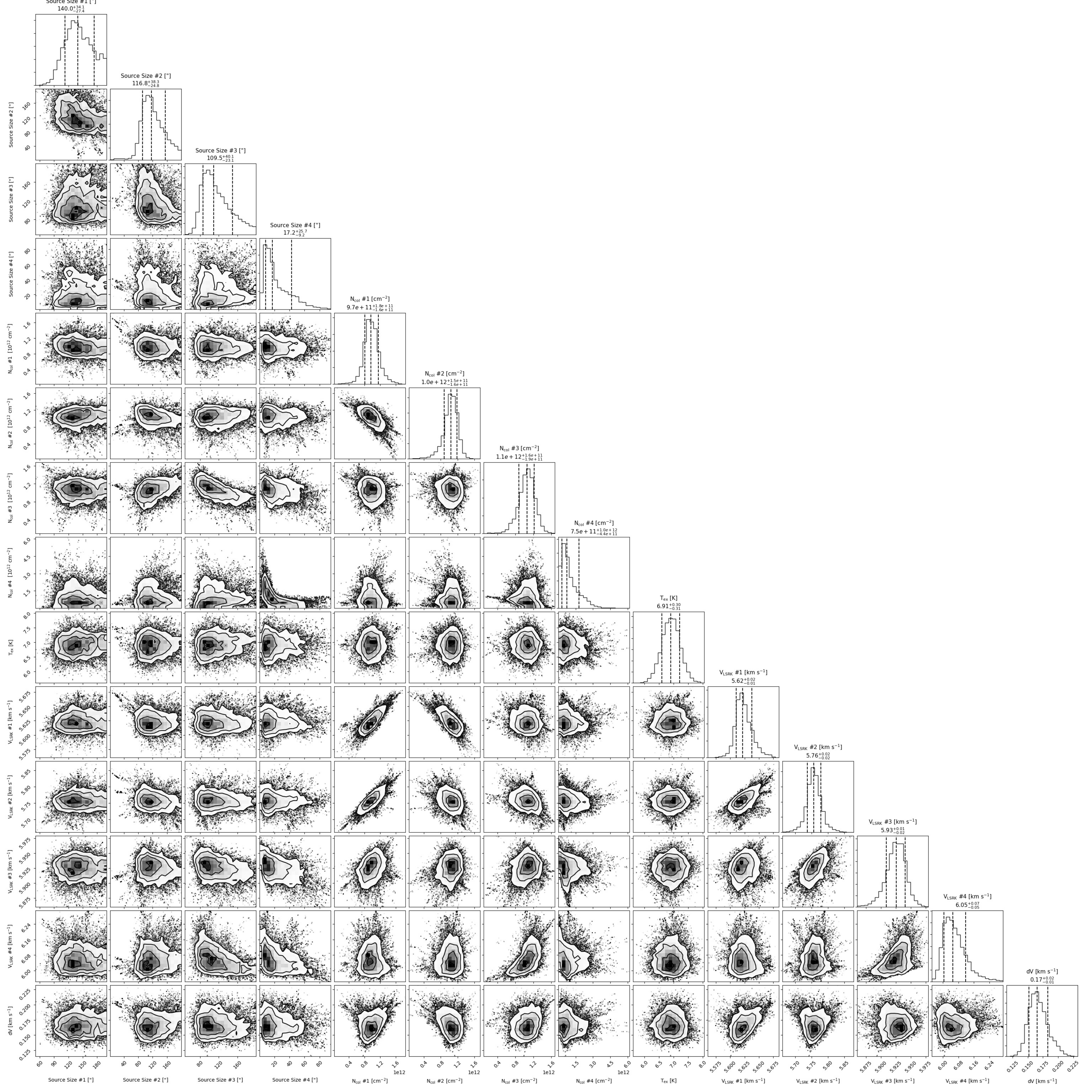}
\caption{Parameter covariances and marginalized posterior distributions for the \ce{HCCNC} MCMC fit. 16$^{th}$, 50$^{th}$, and 84$^{th}$ confidence intervals (corresponding to $\pm$1 sigma for a Gaussian posterior distribution) are shown as vertical lines. }
\label{hc2nc_triangle}
\end{figure*}

{Of these transitions, 6 transitions are above the 5\% threshold, which was uniformly applied to the whole GOTHAM dataset, and were therefore considered for the MCMC fitting and spectral stacking process, the data used in which are available in \citet{GOTHAMDR1}.} The resulting best-fit parameters are given in Table~\ref{hc2nc_results}. {The noise level of the ${1_0\ \rightarrow\ 0_1}$ spectrum is ${\sim}3$ mK, which accounts for the apparent difference seen between the constructed and observed profiles.} The stacked spectrum and matched filter results are shown in Figure~\ref{hc2nc_stack}, while a corner plot of the parameter covariances for the \ce{HCCNC} MCMC fit is shown in Figure~\ref{hc2nc_triangle}.

\begin{deluxetable}{ccrccr}[hbt!]
    \tablecaption{Spectroscopic Properties of the \ce{HCCNC} lines \label{tab:spc-par-hc2nc}}
    \tablewidth{\columnwidth}
    \tablehead{
    \multicolumn{2}{c}{Transitions} &\colhead{Frequency}   &\colhead{$E_{up}$}   &\colhead{$\log_{10}{\frac{A_{ul}}{\mathrm{s^{-1}}}}$}   &\colhead{$S_{ij}\mu^{2}$}\\
    \colhead{$J'\ \rightarrow\ J''$}&\colhead{$F'\ \rightarrow\ F''$}&\colhead{(MHz)}&\colhead{(K)}&&\colhead{(Debye$^2$)}\\}
    \startdata
$1\ \rightarrow\ 0$& $0\ \rightarrow\ 1$&        9935.2000(150) &0.48 &  -7.4859&       2.86 \\ 
$1\ \rightarrow\ 0$& $2\ \rightarrow\ 1$&        9935.6270(150) &0.48 &  -7.4859&       14.31\\ 
$1\ \rightarrow\ 0$& $1\ \rightarrow\ 1$&        9935.9100(150) &0.48 &  -7.4858&       8.59 \\ 
$3\ \rightarrow\ 2$& $2\ \rightarrow\ 2$&       29806.5354(122) &2.86 &  -6.7535&       2.86 \\ 
& $2\ \rightarrow\ 3$&       29806.8398(39)  &2.86 &  -8.2976&       0.08 \\ 
& $4\ \rightarrow\ 3$&       29806.9503(20)  &2.86 &  -5.9454&       33.11\\ 
& $3\ \rightarrow\ 2$&       29806.9615(20)  &2.86 &  -5.9965&       22.89\\ 
& $2\ \rightarrow\ 1$&       29807.0089(25)  &2.86 &  -6.0211&       15.45\\ 
& $3\ \rightarrow\ 3$&       29807.2660(89)  &2.86 &  -6.8996&       2.86
    \enddata
    \tablecomments{The spectroscopic data of the \ce{HCCNC} transitions corresponding to the {six} detected lines are taken from the JPL catalogue\footnote{\footnotesize{https://spec.jpl.nasa.gov}} and the SPLATALOGUE spectroscopy database, which are based on the FTMW and millimetre-wave measurements of \citet{1992JMoSp.156...39G} and \citet{1993JMoSp.158..298K}.}
\end{deluxetable}

\begin{table*}[hbt!] 
\centering
\caption{\ce{HCCNC} best-fit parameters from the MCMC analysis}
\begin{tabular}{c c c c c c}
\toprule
\multirow{2}{*}{Component}&	$v_{lsr}$					&	Size					&	\multicolumn{1}{c}{$N_T^\dagger$}					&	$T_{ex}$							&	$\Delta V$		\\
			&	(km s$^{-1}$)				&	($^{\prime\prime}$)		&	\multicolumn{1}{c}{(10$^{12}$ cm$^{-2}$)}		&	(K)								&	(km s$^{-1}$)	\\
\midrule
\hspace{0.1em}\vspace{-0.5em}\\
C1	&	$5.622^{+0.016}_{-0.011}$	&	$140^{+34}_{-27}$	&	$0.97^{+0.18}_{-0.16}$	&	\multirow{6}{*}{$6.9^{+0.3}_{-0.3}$}	&	\multirow{6}{*}{$0.166^{+0.017}_{-0.014}$}\\
\hspace{0.1em}\vspace{-0.5em}\\
C2	&	$5.756^{+0.022}_{-0.020}$	&	$117^{+38}_{-25}$	&	$1.04^{+0.15}_{-0.16}$	&		&	\\
\hspace{0.1em}\vspace{-0.5em}\\
C3	&	$5.926^{+0.015}_{-0.017}$	&	$110^{+40}_{-23}$	&	$1.06^{+0.16}_{-0.19}$	&		&	\\
\hspace{0.1em}\vspace{-0.5em}\\
C4	&	$6.051^{+0.066}_{-0.045}$	&	$17^{+26}_{-9}$	&	$0.75^{+1.02}_{-0.44}$	&		&	\\
\hspace{0.1em}\vspace{-0.5em}\\
\midrule
$N_T$ (Total)$^{\dagger\dagger}$	&	 \multicolumn{5}{c}{$3.82^{+1.06}_{-0.53}\times 10^{12}$~cm$^{-2}$}\\
\bottomrule
\end{tabular}

\begin{minipage}{0.75\textwidth}
	\footnotesize
	\textbf{Note} -- The quoted uncertainties represent the 16$^{th}$ and 84$^{th}$ percentile ($1\sigma$ for a Gaussian distribution) uncertainties, which are derived with the same methods mentioned in Table~\ref{hc4nc_results}. See Figure~\ref{hc2nc_triangle} for a covariance plot.
\end{minipage}

\label{hc2nc_results}

\end{table*}

\section{\texorpdfstring{\ce{HC6NC}}{HC6NC} Analysis Results \label{apx:HC6NC}}
\restartappendixnumbering
%{A total of 10 transitions (including hyperfine components) of \ce{HC6NC} were covered by the observations and were above our predicted flux threshold of 5\%, as discussed in \citet{Loomis:2020aa}. Of these transitions, none were coincident with interfering transitions of other species, and thus a total of 10 transitions were therefore considered \citep{GOTHAMDR1}.} In our observation, no signal beyond a $1\sigma$ detection limit can be assigned to \ce{HC6NC}. Column density upper limits are therefore constrained using the modified fitting process described in \citet{Loomis:2020aa}, the results of which are given in Table~\ref{hc6nc_results}. A corner plot of the parameter covariances for the \ce{HC6NC} MCMC fit is shown in Figure~\ref{hc6nc_triangle}.

\begin{figure*}
\centering
\includegraphics[width=\textwidth]{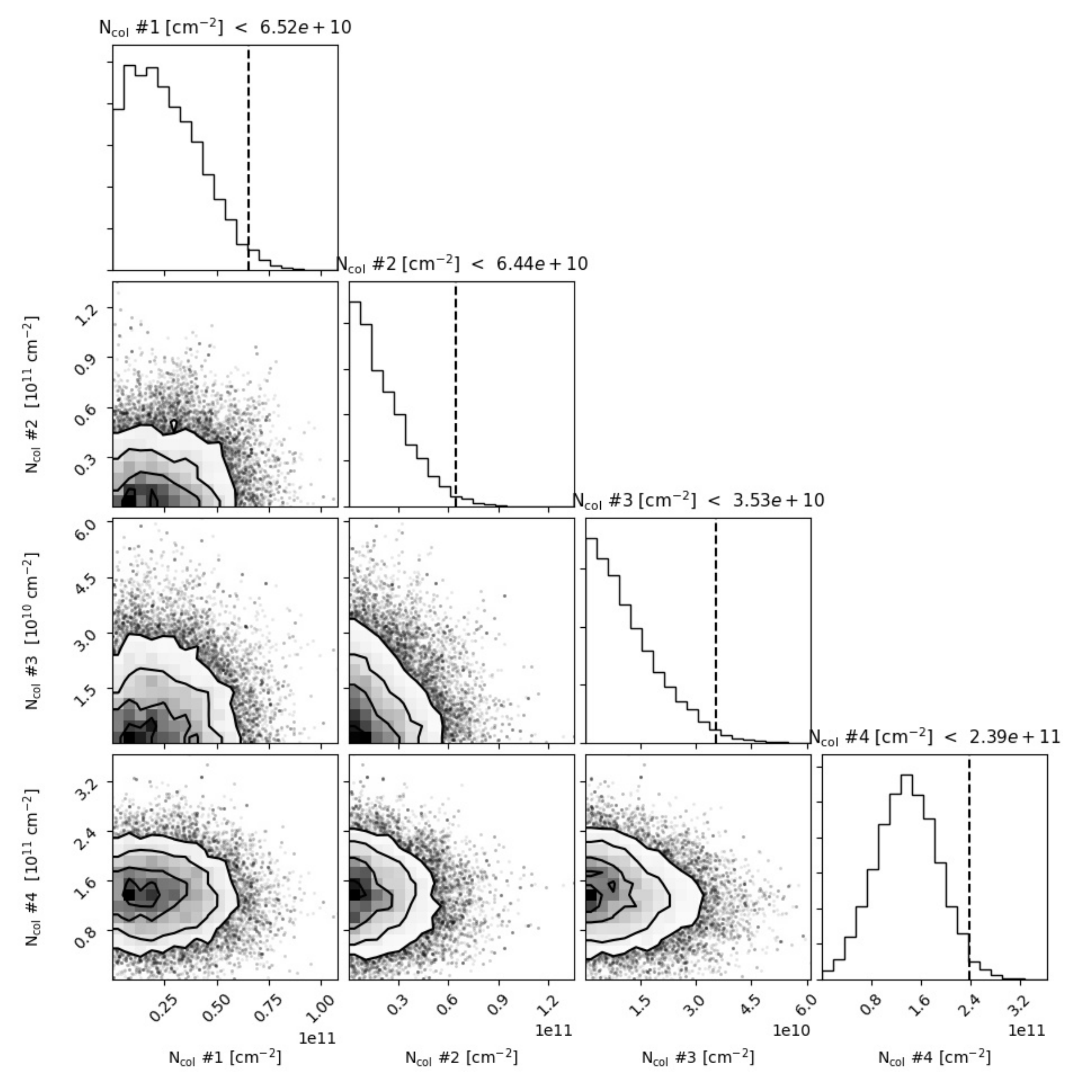}
\caption{Parameter covariances and marginalized posterior distributions for the \ce{HC6NC} MCMC fit. The 97.8$^{th}$ confidence interval (corresponding to 2 sigmas for a Gaussian posterior distribution) is shown as a vertical line.}
\label{hc6nc_triangle}
\end{figure*}

{Following the similar line-selection process with \ce{HC4NC}, a total of 10 transitions (including hyperfine components) of \ce{HC6NC} were considered and the data are again available in \citet{GOTHAMDR1}.} In our observation, no signal beyond a $1\sigma$ detection limit can be assigned to \ce{HC6NC}. Column density upper limits are therefore constrained using the modified fitting process described in \citet{Loomis:2020aa}, the results of which are given in Table~\ref{hc6nc_results}. A corner plot of the parameter covariances for the \ce{HC6NC} MCMC fit is shown in Figure~\ref{hc6nc_triangle}.

\begin{table*}[hbt!]
\centering
\caption{\ce{HC6NC} derived upper limit column densities from the MCMC analysis}
\begin{tabular}{c c c c c c }
\toprule
\multirow{2}{*}{Component}&	$v_{lsr}$					&	Size					&	\multicolumn{1}{c}{$N_T^\dagger$}					&	$T_{ex}$							&	$\Delta V$		\\
			&	(km s$^{-1}$)				&	($^{\prime\prime}$)		&	\multicolumn{1}{c}{(10$^{11}$ cm$^{-2}$)}		&	(K)								&	(km s$^{-1}$)	\\
\midrule
\hspace{0.1em}\vspace{-0.5em}\\
C1	&	[$5.624$]	&	[$33$]	&	$<$$0.65$	&	\multirow{6}{*}{[$6.5$]}	&	\multirow{6}{*}{[$0.117$]}\\
\hspace{0.1em}\vspace{-0.5em}\\
C2	&	[$5.790$]	&	[$22$]	&	$<$$0.64$	&		&	\\
\hspace{0.1em}\vspace{-0.5em}\\
C3	&	[$5.910$]	&	[$50$]	&	$<$$0.35$	&		&	\\
\hspace{0.1em}\vspace{-0.5em}\\
C4	&	[$6.033$]	&	[$18$]	&	$<$$2.39$	&		&	\\
\hspace{0.1em}\vspace{-0.5em}\\
\midrule
$N_T$ (Total)$^{\dagger\dagger}$	&	 \multicolumn{5}{c}{$<4.04\times 10^{11}$~cm$^{-2}$}\\
\bottomrule
\end{tabular}

\begin{minipage}{0.75\textwidth}
	\footnotesize
	\textbf{Note} -- Upper limits are given as the 97.8$^{th}$ percentile (2$\sigma$) value. Parameters in brackets were held fixed to the 50$^{th}$ percentile value. See Figure~\ref{hc6nc_triangle} for a covariance plot.
\end{minipage}

\label{hc6nc_results}

\end{table*}

\begin{thebibliography}{}
\expandafter\ifx\csname natexlab\endcsname\relax\def\natexlab#1{#1}\fi
\providecommand{\url}[1]{\href{#1}{#1}}
\providecommand{\dodoi}[1]{doi:~\href{http://doi.org/#1}{\nolinkurl{#1}}}
\providecommand{\doeprint}[1]{\href{http://ascl.net/#1}{\nolinkurl{http://ascl.net/#1}}}
\providecommand{\doarXiv}[1]{\href{https://arxiv.org/abs/#1}{\nolinkurl{https://arxiv.org/abs/#1}}}

\bibitem[{{Acharyya} \& {Herbst}(2017)}]{2017ApJ...850..105A}
{Acharyya}, K., \& {Herbst}, E. 2017, \apj, 850, 105,
  \dodoi{10.3847/1538-4357/aa937e}

\bibitem[{{Araya} {et~al.}(2005){Araya}, {Hofner}, {Kurtz}, {Bronfman}, \&
  {DeDeo}}]{2005ApJS..157..279A}
{Araya}, E., {Hofner}, P., {Kurtz}, S., {Bronfman}, L., \& {DeDeo}, S. 2005,
  \apjs, 157, 279, \dodoi{10.1086/427187}

\bibitem[{{Balucani} {et~al.}(2015){Balucani}, {Ceccarelli}, \&
  {Taquet}}]{2015MNRAS.449L..16B}
{Balucani}, N., {Ceccarelli}, C., \& {Taquet}, V. 2015, \mnras, 449, L16,
  \dodoi{10.1093/mnrasl/slv009}

\bibitem[{{Bell} {et~al.}(1998){Bell}, {Watson}, {Feldman}, \&
  {Travers}}]{1998ApJ...508..286B}
{Bell}, M.~B., {Watson}, J.~K.~G., {Feldman}, P.~A., \& {Travers}, M.~J. 1998,
  \apj, 508, 286, \dodoi{10.1086/306405}

\bibitem[{{Botschwina} {et~al.}(1998){Botschwina}, {Heyl}, {Chen}, {McCarthy},
  {Grabow}, {Travers}, \& {Thaddeus}}]{1998JChPh.109.3108B}
{Botschwina}, P., {Heyl}, {\"A}., {Chen}, W., {et~al.} 1998, \jcp, 109, 3108,
  \dodoi{10.1063/1.476515}

\bibitem[{Botschwina {et~al.}(1993)Botschwina, Horn, FlÃŒgge, \&
  Seeger}]{FT9938902219}
Botschwina, P., Horn, M., FlÃŒgge, J., \& Seeger, S. 1993, J. Chem. Soc.{,}
  Faraday Trans., 89, 2219, \dodoi{10.1039/FT9938902219}

\bibitem[{{Broten} {et~al.}(1978){Broten}, {Oka}, {Avery}, {MacLeod}, \&
  {Kroto}}]{1978ApJ...223L.105B}
{Broten}, N.~W., {Oka}, T., {Avery}, L.~W., {MacLeod}, J.~M., \& {Kroto}, H.~W.
  1978, \apjl, 223, L105, \dodoi{10.1086/182739}

\bibitem[{{Brown}(1977)}]{1977Natur.270...39B}
{Brown}, R.~D. 1977, \nat, 270, 39, \dodoi{10.1038/270039a0}

\bibitem[{{Brown} {et~al.}(1989){Brown}, {Burden}, \&
  {Cuno}}]{1989ApJ...347..855B}
{Brown}, R.~D., {Burden}, F.~R., \& {Cuno}, A. 1989, \apj, 347, 855,
  \dodoi{10.1086/168175}

\bibitem[{{Burkhardt} {et~al.}(2018){Burkhardt}, {Herbst}, {Kalenskii},
  {McCarthy}, {Remijan}, \& {McGuire}}]{2018MNRAS.474.5068B}
{Burkhardt}, A.~M., {Herbst}, E., {Kalenskii}, S.~V., {et~al.} 2018, \mnras,
  474, 5068, \dodoi{10.1093/mnras/stx2972}

\bibitem[{Burkhardt {et~al.}(2020)Burkhardt, Loomis, Shingledecker, Lee,
  Remijan, McCarthy, \& McGuire}]{Burkhardt:2020aa}
Burkhardt, A.~M., Loomis, R.~A., Shingledecker, C.~N., {et~al.} 2020, Nature
  Astronomy, submitted

\bibitem[{{Calcutt} {et~al.}(2018){Calcutt}, {Fiechter}, {Willis},
  {M{\"u}ller}, {Garrod}, {J{\o}rgensen}, {Wampfler}, {Bourke}, {Coutens},
  {Drozdovskaya}, {Ligterink}, \& {Kristensen}}]{2018AnA...617A..95C}
{Calcutt}, H., {Fiechter}, M.~R., {Willis}, E.~R., {et~al.} 2018, \aap, 617,
  A95, \dodoi{10.1051/0004-6361/201833140}

\bibitem[{{Coutens} {et~al.}(2017){Coutens}, {Rawlings}, {Viti}, \&
  {Williams}}]{2017MNRAS.467..737C}
{Coutens}, A., {Rawlings}, J.~M.~C., {Viti}, S., \& {Williams}, D.~A. 2017,
  \mnras, 467, 737, \dodoi{10.1093/mnras/stx119}

\bibitem[{{Dobashi} {et~al.}(2018){Dobashi}, {Shimoikura}, {Nakamura},
  {Kameno}, {Mizuno}, \& {Taniguchi}}]{2018ApJ...864...82D}
{Dobashi}, K., {Shimoikura}, T., {Nakamura}, F., {et~al.} 2018, \apj, 864, 82,
  \dodoi{10.3847/1538-4357/aad62f}

\bibitem[{{Gensheimer}(1997)}]{1997ApJ...479L..75G}
{Gensheimer}, P.~D. 1997, \apjl, 479, L75, \dodoi{10.1086/310576}

\bibitem[{{Geppert} {et~al.}(2004){Geppert}, {Ehlerding}, {Hellberg},
  {Semaniak}, {{\"O}sterdahl}, {Kami{\'n}ska}, {Al-Khalili}, {Zhaunerchyk},
  {Thomas}, \& {af Ugglas}}]{2004ApJ...613.1302G}
{Geppert}, W.~D., {Ehlerding}, A., {Hellberg}, F., {et~al.} 2004, \apj, 613,
  1302, \dodoi{10.1086/422335}

\bibitem[{{GOTHAM Collaboration}(2020)}]{GOTHAMDR1}
{GOTHAM Collaboration}. 2020, {Spectral Stacking Data for Phase 1 Science
  Release of GOTHAM}, 4.0,  Harvard Dataverse, \dodoi{10.7910/DVN/PG7BHO}.
\newblock \url{https://doi.org/10.7910/DVN/PG7BHO}

\bibitem[{{Graninger} {et~al.}(2015){Graninger}, {{\"O}berg}, {Qi}, \&
  {Kastner}}]{2015ApJ...807L..15G}
{Graninger}, D., {{\"O}berg}, K.~I., {Qi}, C., \& {Kastner}, J. 2015, \apjl,
  807, L15, \dodoi{10.1088/2041-8205/807/1/L15}

\bibitem[{{Graninger} {et~al.}(2014){Graninger}, {Herbst}, {{\"O}berg}, \&
  {Vasyunin}}]{2014ApJ...787...74G}
{Graninger}, D.~M., {Herbst}, E., {{\"O}berg}, K.~I., \& {Vasyunin}, A.~I.
  2014, \apj, 787, 74, \dodoi{10.1088/0004-637X/787/1/74}

\bibitem[{{Gratier} {et~al.}(2016){Gratier}, {Majumdar}, {Ohishi}, {Roueff},
  {Loison}, {Hickson}, \& {Wakelam}}]{2016ApJS..225...25G}
{Gratier}, P., {Majumdar}, L., {Ohishi}, M., {et~al.} 2016, \apjs, 225, 25,
  \dodoi{10.3847/0067-0049/225/2/25}

\bibitem[{{Gratier} {et~al.}(2013){Gratier}, {Pety}, {Guzm{\'a}n}, {Gerin},
  {Goicoechea}, {Roueff}, \& {Faure}}]{2013A&A...557A.101G}
{Gratier}, P., {Pety}, J., {Guzm{\'a}n}, V., {et~al.} 2013, \aap, 557, A101,
  \dodoi{10.1051/0004-6361/201321031}

\bibitem[{{Gronowski} \& {Ko{\l}os}(2006)}]{2006CPL...428..245G}
{Gronowski}, M., \& {Ko{\l}os}, R. 2006, Chemical Physics Letters, 428, 245,
  \dodoi{10.1016/j.cplett.2006.07.041}

\bibitem[{{Guarnieri} {et~al.}(1992){Guarnieri}, {Hinze}, {Kr{\"u}ger},
  {Zerbe-Foese}, {Lentz}, \& {Preugschat}}]{1992JMoSp.156...39G}
{Guarnieri}, A., {Hinze}, R., {Kr{\"u}ger}, M., {et~al.} 1992, Journal of
  Molecular Spectroscopy, 156, 39, \dodoi{10.1016/0022-2852(92)90091-2}

\bibitem[{{Hacar} {et~al.}(2020){Hacar}, {Bosman}, \& {van
  Dishoeck}}]{2020A&A...635A...4H}
{Hacar}, A., {Bosman}, A.~D., \& {van Dishoeck}, E.~F. 2020, \aap, 635, A4,
  \dodoi{10.1051/0004-6361/201936516}

\bibitem[{{Haykal} {et~al.}(2013){Haykal}, {Margul{\`e}s}, {Huet}, {Motyienko},
  {{\'E}cija}, {Cocinero}, {Basterretxea}, {Fern{\'a}ndez}, {Casta{\~n}o},
  {Lesarri}, {Guillemin}, {Tercero}, \& {Cernicharo}}]{2013ApJ...777..120H}
{Haykal}, I., {Margul{\`e}s}, L., {Huet}, T.~R., {et~al.} 2013, \apj, 777, 120,
  \dodoi{10.1088/0004-637X/777/2/120}

\bibitem[{{Herbst} {et~al.}(2000){Herbst}, {Terzieva}, \&
  {Talbi}}]{2000MNRAS.311..869H}
{Herbst}, E., {Terzieva}, R., \& {Talbi}, D. 2000, \mnras, 311, 869,
  \dodoi{10.1046/j.1365-8711.2000.03103.x}

\bibitem[{{Hincelin} {et~al.}(2011){Hincelin}, {Wakelam}, {Hersant},
  {Guilloteau}, {Loison}, {Honvault}, \& {Troe}}]{2011A&A...530A..61H}
{Hincelin}, U., {Wakelam}, V., {Hersant}, F., {et~al.} 2011, \aap, 530, A61,
  \dodoi{10.1051/0004-6361/201016328}

\bibitem[{{Hirota} {et~al.}(1998){Hirota}, {Yamamoto}, {Mikami}, \&
  {Ohishi}}]{1998ApJ...503..717H}
{Hirota}, T., {Yamamoto}, S., {Mikami}, H., \& {Ohishi}, M. 1998, \apj, 503,
  717, \dodoi{10.1086/306032}

\bibitem[{{Hung} {et~al.}(2019){Hung}, {Liu}, {Su}, {He}, {Lee}, {Takahashi},
  \& {Chen}}]{2019ApJ...872...61H}
{Hung}, T., {Liu}, S.-Y., {Su}, Y.-N., {et~al.} 2019, \apj, 872, 61,
  \dodoi{10.3847/1538-4357/aafc23}

\bibitem[{{Irvine} \& {Schloerb}(1984)}]{1984ApJ...282..516I}
{Irvine}, W.~M., \& {Schloerb}, F.~P. 1984, \apj, 282, 516,
  \dodoi{10.1086/162229}

\bibitem[{{Kawaguchi} {et~al.}(1992){Kawaguchi}, {Ohishi}, {Ishikawa}, \&
  {Kaifu}}]{1992ApJ...386L..51K}
{Kawaguchi}, K., {Ohishi}, M., {Ishikawa}, S.-I., \& {Kaifu}, N. 1992, \apjl,
  386, L51, \dodoi{10.1086/186290}

\bibitem[{{Kruger} {et~al.}(1993){Kruger}, {Stahl}, \&
  {Dreizler}}]{1993JMoSp.158..298K}
{Kruger}, M., {Stahl}, W., \& {Dreizler}, H. 1993, Journal of Molecular
  Spectroscopy, 158, 298, \dodoi{10.1006/jmsp.1993.1074}

\bibitem[{{Larsson} {et~al.}(2012){Larsson}, {Geppert}, \&
  {Nyman}}]{2012RPPh...75f6901L}
{Larsson}, M., {Geppert}, W.~D., \& {Nyman}, G. 2012, Reports on Progress in
  Physics, 75, 066901, \dodoi{10.1088/0034-4885/75/6/066901}

\bibitem[{{Little} {et~al.}(1978){Little}, {MacDonald}, {Riley}, \&
  {Matheson}}]{1978MNRAS.183P..45L}
{Little}, L.~T., {MacDonald}, G.~H., {Riley}, P.~W., \& {Matheson}, D.~N. 1978,
  \mnras, 183, 45P, \dodoi{10.1093/mnras/183.1.45P}

\bibitem[{{Loison} {et~al.}(2014{\natexlab{a}}){Loison}, {Wakelam}, \&
  {Hickson}}]{2014MNRAS.443..398L}
{Loison}, J.-C., {Wakelam}, V., \& {Hickson}, K.~M. 2014{\natexlab{a}}, \mnras,
  443, 398, \dodoi{10.1093/mnras/stu1089}

\bibitem[{{Loison} {et~al.}(2014{\natexlab{b}}){Loison}, {Wakelam}, {Hickson},
  {Bergeat}, \& {Mereau}}]{2014MNRAS.437..930L}
{Loison}, J.-C., {Wakelam}, V., {Hickson}, K.~M., {Bergeat}, A., \& {Mereau},
  R. 2014{\natexlab{b}}, \mnras, 437, 930, \dodoi{10.1093/mnras/stt1956}

\bibitem[{Loomis {et~al.}(2020)Loomis, Burkhardt, Shingledecker, Charnley,
  Cordiner, Herbst, Kalenskii, Lee, Willis, Xue, Remijan, McCarthy, \&
  McGuire}]{Loomis:2020aa}
Loomis, R.~A., Burkhardt, A.~M., Shingledecker, C.~N., {et~al.} 2020, Nature
  Astronomy, submitted

\bibitem[{{L{\'o}pez} {et~al.}(2014){L{\'o}pez}, {Tercero}, {Kisiel}, {Daly},
  {Berm{\'u}dez}, {Calcutt}, {Marcelino}, {Viti}, {Drouin}, {Medvedev},
  {Neese}, {Pszcz{\'o}{\l}kowski}, {Alonso}, \&
  {Cernicharo}}]{2014A&A...572A..44L}
{L{\'o}pez}, A., {Tercero}, B., {Kisiel}, Z., {et~al.} 2014, \aap, 572, A44,
  \dodoi{10.1051/0004-6361/201423622}

\bibitem[{{Margul{\`e}s} {et~al.}(2018){Margul{\`e}s}, {Tercero}, {Guillemin},
  {Motiyenko}, \& {Cernicharo}}]{2018A&A...610A..44M}
{Margul{\`e}s}, L., {Tercero}, B., {Guillemin}, J.~C., {Motiyenko}, R.~A., \&
  {Cernicharo}, J. 2018, \aap, 610, A44, \dodoi{10.1051/0004-6361/201731515}

\bibitem[{McCarthy {et~al.}(2020)McCarthy, Lee, Loomis, Burkhardt,
  Shingledecker, Charnley, Cordiner, Herbst, Kalenskii, Willis, Xue, Remijan,
  \& McGuire}]{McCarthy:2020aa}
McCarthy, M.~C., Lee, K. L.~K., Loomis, R.~A., {et~al.} 2020, Nature Astronomy,
  submitted

\bibitem[{{McGuire}(2018)}]{2018ApJS..239...17M}
{McGuire}, B.~A. 2018, \apjs, 239, 17, \dodoi{10.3847/1538-4365/aae5d2}

\bibitem[{{McGuire} {et~al.}(2018){McGuire}, {Burkhardt}, {Kalenskii},
  {Shingledecker}, {Remijan}, {Herbst}, \& {McCarthy}}]{2018Sci...359..202M}
{McGuire}, B.~A., {Burkhardt}, A.~M., {Kalenskii}, S., {et~al.} 2018, Science,
  359, 202, \dodoi{10.1126/science.aao4890}

\bibitem[{{McGuire} {et~al.}(2017){McGuire}, {Burkhardt}, {Shingledecker},
  {Kalenskii}, {Herbst}, {Remijan}, \& {McCarthy}}]{2017ApJ...843L..28M}
{McGuire}, B.~A., {Burkhardt}, A.~M., {Shingledecker}, C.~N., {et~al.} 2017,
  \apjl, 843, L28, \dodoi{10.3847/2041-8213/aa7ca3}

\bibitem[{McGuire {et~al.}(2020{\natexlab{a}})McGuire, Burkhardt, Loomis,
  Shingledecker, Lee, Charnley, Cordiner, Herbst, Kalenskii, Momjian, Willis,
  Xue, Remijan, \& McCarthy}]{McGuire:2020bb}
McGuire, B.~A., Burkhardt, A.~M., Loomis, R.~A., {et~al.} 2020{\natexlab{a}},
  Astrophysical Journal Letters, submitted

\bibitem[{McGuire {et~al.}(2020{\natexlab{b}})McGuire, Loomis, Burkhardt, Lee,
  Shingledecker, Charnley, Cordiner, Herbst, Kalenskii, Willis, Xue, Remijan,
  \& McCarthy}]{McGuire:2020aa}
McGuire, B.~A., Loomis, R.~A., Burkhardt, A.~M., {et~al.} 2020{\natexlab{b}},
  Science, submitted

\bibitem[{{Miao} \& {Snyder}(1997)}]{1997ApJ...480L..67M}
{Miao}, Y., \& {Snyder}, L.~E. 1997, \apjl, 480, L67, \dodoi{10.1086/310624}

\bibitem[{{M{\"u}ller} {et~al.}(2005){M{\"u}ller}, {Schl{\"o}der}, {Stutzki},
  \& {Winnewisser}}]{2005JMoSt.742..215M}
{M{\"u}ller}, H. S.~P., {Schl{\"o}der}, F., {Stutzki}, J., \& {Winnewisser}, G.
  2005, Journal of Molecular Structure, 742, 215,
  \dodoi{10.1016/j.molstruc.2005.01.027}

\bibitem[{{Osamura} {et~al.}(1999){Osamura}, {Fukuzawa}, {Terzieva}, \&
  {Herbst}}]{1999ApJ...519..697O}
{Osamura}, Y., {Fukuzawa}, K., {Terzieva}, R., \& {Herbst}, E. 1999, \apj, 519,
  697, \dodoi{10.1086/307406}

\bibitem[{{Qu{\'e}nard} {et~al.}(2017){Qu{\'e}nard}, {Vastel}, {Ceccarelli},
  {Hily-Blant}, {Lefloch}, \& {Bachiller}}]{2017MNRAS.470.3194Q}
{Qu{\'e}nard}, D., {Vastel}, C., {Ceccarelli}, C., {et~al.} 2017, \mnras, 470,
  3194, \dodoi{10.1093/mnras/stx1373}

\bibitem[{{Remijan} {et~al.}(2005){Remijan}, {Hollis}, {Lovas}, {Plusquellic},
  \& {Jewell}}]{2005ApJ...632..333R}
{Remijan}, A.~J., {Hollis}, J.~M., {Lovas}, F.~J., {Plusquellic}, D.~F., \&
  {Jewell}, P.~R. 2005, \apj, 632, 333, \dodoi{10.1086/432908}

\bibitem[{{Ruaud} {et~al.}(2016){Ruaud}, {Wakelam}, \&
  {Hersant}}]{2016MNRAS.459.3756R}
{Ruaud}, M., {Wakelam}, V., \& {Hersant}, F. 2016, \mnras, 459, 3756,
  \dodoi{10.1093/mnras/stw887}

\bibitem[{{Schilke} {et~al.}(1992){Schilke}, {Walmsley}, {Pineau Des Forets},
  {Roueff}, {Flower}, \& {Guilloteau}}]{1992A&A...256..595S}
{Schilke}, P., {Walmsley}, C.~M., {Pineau Des Forets}, G., {et~al.} 1992, \aap,
  256, 595

\bibitem[{Stanton {et~al.}(2017)Stanton, Gauss, Cheng, Harding, Matthews, \&
  Szalay}]{stanton_cfour_2017}
Stanton, J.~F., Gauss, J., Cheng, L., {et~al.} 2017, {CFOUR},
  {Coupled}-{Cluster} techniques for {Computational} {Chemistry}, a
  quantum-chemical program package

\bibitem[{{Takagi} {et~al.}(1999){Takagi}, {Fukuzawa}, {Osamura}, \&
  {Schaefer}}]{1999ApJ...525..791T}
{Takagi}, N., {Fukuzawa}, K., {Osamura}, Y., \& {Schaefer}, Henry~F., I. 1999,
  \apj, 525, 791, \dodoi{10.1086/307914}

\bibitem[{{Takano} {et~al.}(1998){Takano}, {Masuda}, {Hirahara}, {Suzuki},
  {Ohishi}, {Ishikawa}, {Kaifu}, {Kasai}, {Kawaguchi}, \&
  {Wilson}}]{1998A&A...329.1156T}
{Takano}, S., {Masuda}, A., {Hirahara}, Y., {et~al.} 1998, \aap, 329, 1156

\bibitem[{{Taniguchi} {et~al.}(2016){Taniguchi}, {Ozeki}, {Saito}, {Sakai},
  {Nakamura}, {Kameno}, {Takano}, \& {Yamamoto}}]{2016ApJ...817..147T}
{Taniguchi}, K., {Ozeki}, H., {Saito}, M., {et~al.} 2016, \apj, 817, 147,
  \dodoi{10.3847/0004-637X/817/2/147}

\bibitem[{{Tennekes} {et~al.}(2006){Tennekes}, {Harju}, {Juvela}, \&
  {T{\'o}th}}]{2006A&A...456.1037T}
{Tennekes}, P.~P., {Harju}, J., {Juvela}, M., \& {T{\'o}th}, L.~V. 2006, \aap,
  456, 1037, \dodoi{10.1051/0004-6361:20040294}

\bibitem[{{Turner} {et~al.}(1997){Turner}, {Pirogov}, \&
  {Minh}}]{1997ApJ...483..235T}
{Turner}, B.~E., {Pirogov}, L., \& {Minh}, Y.~C. 1997, \apj, 483, 235,
  \dodoi{10.1086/304228}

\bibitem[{{Vastel} {et~al.}(2018){Vastel}, {Kawaguchi}, {Qu{\'e}nard},
  {Ohishi}, {Lefloch}, {Bachiller}, \& {M{\"u}ller}}]{2018MNRAS.474L..76V}
{Vastel}, C., {Kawaguchi}, K., {Qu{\'e}nard}, D., {et~al.} 2018, \mnras, 474,
  L76, \dodoi{10.1093/mnrasl/slx197}

\bibitem[{{Vastel} {et~al.}(2019){Vastel}, {Loison}, {Wakelam}, \&
  {Lefloch}}]{2019AnA...625A..91V}
{Vastel}, C., {Loison}, J.~C., {Wakelam}, V., \& {Lefloch}, B. 2019, \aap, 625,
  A91, \dodoi{10.1051/0004-6361/201935010}

\bibitem[{{Vigren} {et~al.}(2012){Vigren}, {Semaniak}, {Hamberg},
  {Zhaunerchyk}, {Kaminska}, {Thomas}, {af Ugglas}, {Larsson}, \&
  {Geppert}}]{2012P&SS...60..102V}
{Vigren}, E., {Semaniak}, J., {Hamberg}, M., {et~al.} 2012, \planss, 60, 102,
  \dodoi{10.1016/j.pss.2011.03.001}

\bibitem[{{Wakelam} {et~al.}(2015){Wakelam}, {Loison}, {Herbst}, {Pavone},
  {Bergeat}, {B{\'e}roff}, {Chabot}, {Faure}, {Galli}, \&
  {Geppert}}]{2015ApJS..217...20W}
{Wakelam}, V., {Loison}, J.~C., {Herbst}, E., {et~al.} 2015, \apjs, 217, 20,
  \dodoi{10.1088/0067-0049/217/2/20}

\bibitem[{{Woon} \& {Herbst}(2009)}]{2009ApJS..185..273W}
{Woon}, D.~E., \& {Herbst}, E. 2009, \apjs, 185, 273,
  \dodoi{10.1088/0067-0049/185/2/273}

\bibitem[{{Xue} {et~al.}(2019){Xue}, {Remijan}, {Burkhardt}, \&
  {Herbst}}]{2019ApJ...871..112X}
{Xue}, C., {Remijan}, A.~J., {Burkhardt}, A.~M., \& {Herbst}, E. 2019, \apj,
  871, 112, \dodoi{10.3847/1538-4357/aaf738}

\bibitem[{Xue {et~al.}(2020)Xue, Willis, Loomis, Lee, Burkhardt, Shingledecker,
  Charnley, Cordiner, Kalenskii, McCarthy, Herbst, Remijan, \&
  McGuire}]{Xue:2020aa}
Xue, C., Willis, E.~R., Loomis, R.~A., {et~al.} 2020, Astrophysical Journal
  Letters, submitted

\bibitem[{{Zeng} {et~al.}(2019){Zeng}, {Qu{\'e}nard}, {Jim{\'e}nez-Serra},
  {Mart{\'\i}n-Pintado}, {Rivilla}, {Testi}, \&
  {Mart{\'\i}n-Dom{\'e}nech}}]{2019MNRAS.484L..43Z}
{Zeng}, S., {Qu{\'e}nard}, D., {Jim{\'e}nez-Serra}, I., {et~al.} 2019, \mnras,
  484, L43, \dodoi{10.1093/mnrasl/slz002}

\end{thebibliography}
\end{document}